\documentclass[fleqn, usenatbib]{mnras}

\usepackage{newtxtext,newtxmath}
\usepackage[T1]{fontenc}
\usepackage{ae,aecompl}

\usepackage{graphicx}	% Including figure files
\usepackage{amsmath}	% Advanced maths commands
\usepackage{amssymb}	% Extra maths symbols

\usepackage{here}

\usepackage{physics}
\usepackage{tikz}
\usetikzlibrary{shapes.geometric, arrows, backgrounds,fit}
\usepackage[capitalize]{cleveref}

\usepackage{todonotes}

\newcommand*{\Table}[1]{Table~\ref{tab:#1}}

\newcommand*{\myminipage}[2]{\begin{minipage}{#1}\centering{#2}\end{minipage}}

\newcommand{\calG}{\mathcal{G}}
\newcommand{\calL}{\mathcal{L}}
\newcommand{\calN}{\mathcal{N}}
\newcommand{\photo}{\text{photo}}
\newcommand{\spec}{\text{spec}}
\newcommand{\betaani}{\beta_\text{ani}}
\newcommand{\thetahalf}{\theta_{1/2}}
\newcommand{\Rhalf}{R_{1/2}}
\newcommand{\Msun}{M_\odot}
\newcommand{\Odds}{\mathrm{Odds}}

\newcommand{\Besancon}{Besan\c{c}on}

\defcitealias{Ichikawa2017}{KI}	
\newcommand{\KI}{\citetalias{Ichikawa2017}}

\title[$J$-factor estimation with the mixture model]{$J$-factor estimation of Draco, Sculptor and Ursa Minor dwarf spheroidal galaxies with the member/foreground mixture model}

\author[S. Horigome et al.]{
Shun-ichi Horigome,$^{1}$\thanks{E-mail: shunichi.horigome@ipmu.jp}
Kohei Hayashi,$^{2}$
Masahiro Ibe,$^{1,2}$
\newauthor
Miho N. Ishigaki,$^{3}$
Shigeki Matsumoto$^{1}$
and 
Hajime Sugai$^{1}$
\\
$^{1}$Kavli Institute for the Physics and Mathematics of the Universe (Kavli IPMU, WPI), The University of Tokyo, Chiba 277-8583, Japan\\
$^{2}$Institute for Cosmic Ray Research (ICRR), The University of Tokyo, Chiba 277-8583, Japan\\
$^{3}$Astronomical Institute, Tohoku University, Aoba-ku, Sendai 980-8578, Japan
}

\date{Accepted XXX. Received YYY; in original form ZZZ}

\pubyear{2019}

\begin{document}
\label{firstpage}
\pagerange{\pageref{firstpage}--\pageref{lastpage}}
\maketitle

\begin{abstract}
Dwarf spheroidal galaxies (dSphs) are promising targets of indirect detection experiments searching for dark matter (DM) at present universe. Toward robust prediction for the amount of signal flux originating in DM annihilation inside dSphs, a precise determination of DM distributions as well as $J$-factors of the dSphs is particularly important. In this work, we estimate those of Draco, Sculptor, and Ursa Minor dSphs by an improved statistical method in which both foreground stars and dSph member stars are simultaneously taken into account. We define the likelihood function of the method as the so-called conditional one to remove sampling bias of observed stellar data. This improved method enables us to estimate DM distributions and $J$-factors of the dSphs directly from observed stellar data contaminated by foreground stars without imposing stringent membership criteria on the measured quantities.
\end{abstract}

\section{Introduction}
The existence of dark matter (DM) in our universe was strongly confirmed by various astrophysical observations such as dynamics of galaxy clusters \citep{Zwicky1933a}, rotation curves of galaxies \citep{Rubin1978,Rubin1980a}, and gravitational lensing \citep{McLaughlin:1998sb, Bradac:2006er, Clowe:2006eq}. The global analysis of cosmic microwave background, large scale structure and supernovae observation data \citep{Ade:2015xua} tells us that DM is responsible for a quarter of the total energy of the present universe; however the microscopic nature of DM is still unknown. Weakly interacting massive particle (WIMP) is an attractive DM candidate, which can naturally explain DM abundance observed today by the well-established freeze-out mechanism. In particular, WIMP with TeV scale mass is well-motivated from the viewpoint of new physics beyond the standard model of particle physics \citep{Moroi2000, Hisano:2006nn, Bhattacherjee2014, Evans2014UniversalityMediation}, and the WIMP is, in fact, intensively studied after the discovery of the Higgs boson at the Large Hadron Collider experiment.

The most promising way to detect the WIMP is the indirect detection searching for signal from DM rich region. Among various targets of the detection, dwarf spheroidal galaxies (dSphs) associated with the Milky Way are ideal ones, as they contain a large amount of DM with small astrophysical backgrounds \citep{Cholis2012, Lefranc2016} and are located, at most, a few hundred kpc away from our solar system. In fact, the gamma-ray search from dSphs excluded a typical WIMP with the mass less than 100\,GeV \citep{Ackermann2015a}. The signal flux of the detection depends not only on the particle nature of DM but also on an astrophysical factor concerning the dSph, namely the $J$-factor:
\begin{equation}
    J(\Delta\Omega) = \qty[\int_{\Delta\Omega}\dd{\Omega}\int_\text{l.o.s.}\dd{l}\rho_\text{DM}^2(l,\,\Omega)]\ .
\end{equation}
Here we define a DM density profile at a distance $l$ and an angle $\Omega$ by $\rho_\text{DM}(l,\Omega)$, which is estimated by the comparison between the observed velocity dispersion curve of dSph member stars and the theoretical prediction on the dispersion curve from dSph stellar kinematics. It is, however, known that several uncertainties are associated with the estimation: spatially dependent anisotropy of the dispersion \citep{Ullio2016}, non-spherical profile \citep{Bonnivard2015a, Hayashi2016}, size of halo truncation \citep{Geringer-Sameth2015}, contamination of binary stars \citep{Koch:2007ye, Simon2007, Mateo2008}, prior bias of Bayesian analysis \citep{Martinez:2009jh}, and foreground contamination \citep{Bonnivard2016, Ichikawa2017, Ichikawa}.

In particular, it is necessary to take care of the foreground contamination even for the case of future observations yielding a  larger amount of observational data, because the number of foreground stars contributing to the contamination increases along the amount of the data.  Various methods are adopted in conventional analyses to remove contaminating stars such as simple sigma-clipping procedure \citep{Coleman2005b}, membership selection based on the expectation-maximisation (EM) algorithm \citep{Walker2009d}, and full Bayesian analysis with the foreground model of a single Gaussian component \citep{Bonnivard2016}.  In \cite[hereafter \KI]{Ichikawa2017,Ichikawa}, we have investigated the effect of the foreground contamination on the $J$-factor estimations of classical and ultra-faint dSphs. One of the goals of our papers was to develop observational strategies and analysis methods for future instruments such as the Prime Focus Spectrograph (PFS) mounted on the 8.2-m Subaru Telescope.PFS is the next-generation spectrograph of the SuMIRe project \citep{2014PASJ...66R...1T, 2015JATIS...1c5001S, Tamura:2016wsg} with a large field of view ($\sim1.38$ degree diameter) and about 2400 fibres, which allows us to observe not only member-like stars but also many foreground stars simultaneously. We have defined a likelihood function based on the mixture model of a dSph member component and three foreground components motivated by the fitting of the \Besancon\ model \citep{Robin:2003}. Using our constructed likelihood function, we have demonstrated that the likelihood function can successfully reproduce the input parameters of mock observational data thanks to large data sets yielded by the large field of view of PFS. By contrast, membership selection based on the EM algorithm can result in biased $J$-factor estimation when the intrinsic velocity dispersion is not flat. Moreover, the selection method sometimes suffers from the foreground contamination effect, because even a few contaminating stars located at the outer region of a dSph make us overestimate the velocity dispersion.

When we apply \KI 's method to the actual data sets, we however need to treat spatial sampling biases of the observed data \citep{Martinez2011}, because the surface density of observed spectroscopic data is not equal to the actual surface density of a dSph due to the sampling. Therefore, in this work, we improve the mixture model likelihood developed in \KI\ to deal with actual stellar data with the sampling biases. By using this improved likelihood function, we obtain the non-biased $J$-factors of Draco, Sculptor, and Ursa Minor dSphs robustly in terms of the foreground contamination.

The organisation of this paper is as follows: In Section \ref{sec:Methods}, we describe our analysis method, which is separated into two parts, photometric and spectroscopic parts. We define likelihood functions for these two parts in Section \ref{sec:photo} and \ref{sec:spec}, respectively. In Section \ref{sec:sampling}, we discuss a sampling algorithm to obtain posterior probability density functions of model parameters as well as the $J$-factor. In Section \ref{sec:DataSource}, we introduce photometric and spectroscopic data sets for each dSph used in our analysis. In Section \ref{sec:crossmatching} and \ref{sec:cmdcut}, we explain the pre-processing for the data introduced in the previous subsection. In Section \ref{sec:Results}, we show our results of parameter estimation and $J$-factor posteriors. In section \ref{sec:discussion}, we discuss the results of our estimation. We summarise our discussion in Section \ref{sec:Conclusions}.

\section{Methods} \label{sec:Methods}

\begin{figure*}
    \newcommand{\minw}{2cm}
    \newcommand{\minh}{0.5cm}
    \tikzset{every node/.style={minimum width=\minw, minimum height=\minh, text centered, draw=black}}
    \tikzstyle{startstop} = [rectangle, rounded corners, fill=red!30]
    \tikzstyle{io} = [trapezium, trapezium left angle=70, trapezium right angle=110, fill=blue!30]
    \tikzstyle{process} = [rectangle, fill=orange!30]
    \tikzstyle{container} = [dashed,inner sep=0.28cm, rounded corners,fill=yellow!20,minimum height=1cm]
    \tikzstyle{arrow} = [thick,->,>=stealth]
    \centering
    \begin{tikzpicture}[node distance=1cm]
    \node[startstop] (start) {Start};
    \node[io,below of=start] (rawdata) {Photometric and spectroscopic samples};
    \node[process,below of=rawdata] (CMDcut) {Colour-magnitude cut};
        \node[io,below of=CMDcut,xshift=-3cm,yshift=-2mm] (cut1) {photometric samples w/ cut};
        \node[process, below of=cut1] (modelselection1) {Stellar model selection};
        \node[io, below of=modelselection1] (likeli1) {$\calL_\photo(\Theta_\photo)$};
        \node[io, below of=modelselection1, xshift=-2.5cm,align=center] (prior1) {$\pi_\photo(\Theta_\photo)$\\(=const.)};
        \node[process, below of=likeli1] (mcmc1) {MCMC};
        \node[io,below of=mcmc1] (post1) {$p_\text{post}(\Theta_\photo)$};
    \begin{scope}[on background layer]
        \coordinate (aux1) at ([yshift=2mm]cut1.north);
        \node[container,fit=(aux1)(cut1)(modelselection1)(likeli1)(prior1)(mcmc1)(post1)] (back1) {};
        \node at (back1.north) [fill=white,draw] {\textbf{(i) Photometry}};
    \end{scope}
    \coordinate[right of=post1,xshift=1cm] (p0);
        \node[io,below of=CMDcut,xshift=+3cm,yshift=-2mm] (cut2) {Spectroscopic samples w/ cut};
        \node[io,right of=cut2,xshift=3cm] (Besancon) {\Besancon\ model mock};
        \node[process,below of=Besancon, text width=4cm] (Ncomponent) {Foreground model selection \\ (Component number estimation)};
        \node[io, below of=cut2] (likeli2) {$\calL_\spec(\Theta_\spec)$};
        \node[io, below of=cut2,xshift=-2.5cm] (prior2) {$\pi_\spec(\Theta_\spec)$};
        \node[process, below of=cut2,yshift=-1cm] (MCMC2) {MCMC};
        \node[io,below of=MCMC2] (post2) {$p_\text{post}(\Theta_\spec)$};
        \node[io,right of=post2,xshift=1.5cm] (Jfactor) {$J$-factor};
    \begin{scope}[on background layer]
        \coordinate (aux2) at ([yshift=2mm]cut2.north);
        \node[container,fit=(aux2)(cut2)(Besancon)(Ncomponent)(likeli2)(prior2)(MCMC2)(post2)] (back2) {};
        \node at (back2.north) [fill=white,draw] {\textbf{(ii) Spectroscopy}};
    \end{scope}
    \node[startstop, below of=Jfactor,xshift=0cm] (end) {End};
    \draw[arrow] (start) -- (rawdata);
    \draw[arrow] (rawdata) -- (CMDcut);
    \draw[arrow] (CMDcut) -- (cut1);
    \draw[arrow] (cut1) -- (modelselection1);
    \draw[arrow] (modelselection1) -- (likeli1);
    \draw[arrow] (likeli1) -- (mcmc1);
    \draw[arrow] (prior1) -- (mcmc1);
    \draw[arrow] (mcmc1) -- (post1);
    \draw (post1) -- (p0);
    \draw[arrow] (p0) -- (prior2);
    \draw[arrow] (CMDcut) -- (cut2);
    \draw[arrow] (cut2) -- (likeli2);
    \draw[arrow] (likeli2) -- (MCMC2);
    \draw[arrow] (prior2) -- (MCMC2);
    \draw[arrow] (MCMC2) -- (post2);
    \draw[arrow] (post2) -- (Jfactor);
    \draw[arrow] (Jfactor) -- (end);
    \draw[arrow] (Besancon) -- (Ncomponent);
    \draw[arrow] (Ncomponent) -- (likeli2);
    %\draw[arrow] (mcmc1) -- (forphoto);
    \end{tikzpicture}
    \caption{\small Flow chart of our analysis method. See Section \ref{sec:Methods} for more details.}
    \label{fig:flowchart}
\end{figure*}
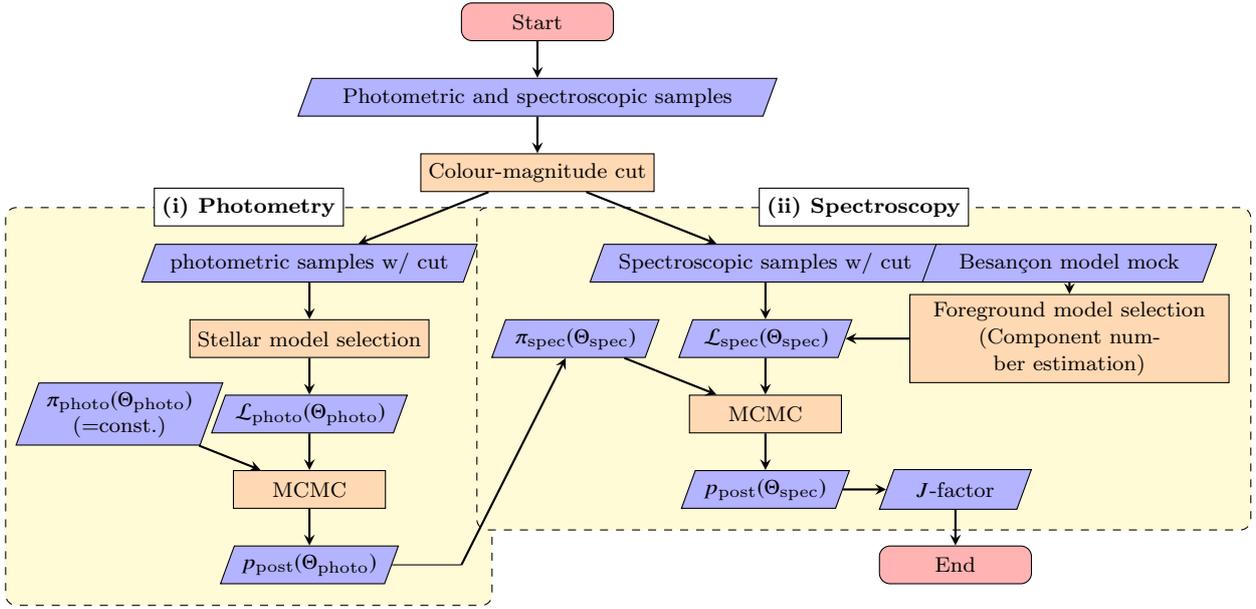

We first show the flowchart of our analysis method in Figure \ref{fig:flowchart}. Our method is mainly separated into two parts as follows:
\begin{enumerate}
\item \textbf{Photometry}: We make use of up-to-date wide-field, multi-band, photometric samples which are available in the public data releases of Sloan Digital Sky Survey \citep{Abolfathi2018}, Pan-STARRS \citep{Chambers2016}, and Dark Energy Survey \citep{Abbott2018}. For each dSph, we impose a colour magnitude cut on the photometric samples to stand out member star candidates on data. With the use of these candidates, we select a suitable stellar density profile from two empirical spatial profiles shown in Eq.(\ref{eq:stellar_profiles}) by comparing their statistical evidence. We then estimate the half-light radius and the local membership probability at this radius.
\item \textbf{Spectroscopy}: We compile the largest samples of stellar line-of-sight velocities which are available from references \citet{Walker2015a}, \citet{Walker2009d}, and \citet{Spencer2018}. Before going to the spectroscopic analysis, we fix the number of independent components of foreground stars by comparing velocity distributions of the data set and of the \Besancon\ model.
Using photometric information (i.e., magnitude of each band) taken from the spectroscopic samples, we impose the same colour-magnitude cut on the spectroscopic samples to avoid sampling bias in terms of colour-magnitudes between the photometric and spectroscopic samples. After that, we estimate the DM halo parameters, and evaluate the posterior distribution of the $J$-factor based on the line-of-sight velocities. We use the posterior parameter distributions of the half-right radius and the local membership probability obtained from the photometric analysis as prior distributions of the spectroscopic analysis.
\end{enumerate}
Here it is worth emphasising that imposing the same criteria on the colour-magnitude for both the photometry and the spectroscopic data means that the membership probability which is tightly fixed to be a specific value by the photometric analysis can be directly applied to the membership probability in the spectroscopic analysis. In addition, we also take the other uncertainty into account in our analysis, which is from the distance between dSph and our solar system. As shown in Appendix \ref{sec:distance_err}, the value of the $J$-factor is scaled as  $J_D \to J_{kD} = k^{-3}J_D$ under the scaling of  $D\to kD$ with $D$ being the distance, and hence it can cause a large uncertainty on the $J$-factor estimation. Note that this scaling of the $J$-factor is always held whenever we adopt the spherical Jean-analysis.

We explain each step of our analysis method in more details in the rest of this section assuming following conditions:
\begin{enumerate}
\item The foreground stars are assumed to be uniformly distributed inside the celestial sphere around the region of interest (RoI).
\item The velocity distribution of the foreground stars is composed of, at most up to three components, which are corresponding to the thin disk, thick disk, and halo components, respectively.
\item The velocity distribution of each foreground component mentioned above is described by a Gaussian distribution.
\item Spectroscopic samples are randomly selected, meaning the samples are selected without respect to their colour-magnitudes.
\item Both the velocity distributions of the member stars and foreground stars do not depend on their colour-magnitudes.
\end{enumerate}
Note again that the number of the components for foreground stars is fixed at the step of the model selection in the spectroscopic part.

\subsection{Photometry}\label{sec:photo}
The purpose of this photometric part is to determine stellar profiles (shapes and half-light radii) of the dSphs and membership probabilities at the half-light radii by using photometric samples, which will be further used in the subsequent spectroscopic analysis.

\subsubsection{Likelihood function}
The half-light radius of the stellar distribution of a dSph and the membership probability at this radius is estimated using the following setup of the likelihood function $\calL_\photo$ and prior $\pi_\photo$:
\begin{align}
\calL_\photo(\Theta_\photo) &= \prod_i 2\pi R_i\qty[s \calN_1\Sigma_1(R_i) + (1-s)\calN_0\Sigma_{0}(R_i)]\ ,\\
\pi_\photo(\Theta_\photo) &: \text{flat.}\ .
\end{align}
Here, $\Theta_\photo$ is the parameter set of the photometry model (to be mentioned later) and $R_i$ denotes the projected radius given by $R_i=D\sin{\theta_i}$, where $\theta_i$ is the separation angle of the $i$-th star from the centre of the dSph distant by $D$ from the Earth. $\Sigma_M(R)$ denotes the projected stellar profile of dSph member stars ($M=1$) and foreground stars ($M=0$). As stated above, we take $\Sigma_0(R) = \text{const.}$ The normalisation factor $\calN_M$ is given by the equation $\calN_M\equiv\qty(\int_0^{R_\photo}\dd{R}2\pi R\Sigma_M(R))^{-1} < 1$ for each $\Sigma_M(R)$, where the integration is performed from the centre ($R=0$) to the maximum radius of the photometric stars denoted by $R_\photo$. See also Table \ref{tab:DataSource}. The coefficient $s$ stands for the global membership probability of the dSph, which is given by the following formula:\footnote{In \KI, the coefficient $s$ is used as one of parameters of the likelihood function even in the spectroscopic analysis. The value of $s$, however, depends on the maximum radius of the photometric samples. Furthermore, the area of the spectroscopic observation is usually not even a circle but a combination of some irregular ones, which complicates the calculation of the global membership probability. See Appendix \ref{App:KI17tomodKI17} for more details.}
\begin{equation}\label{eq:global_and_local_memberships}
    s = \qty[1 + \frac{1}{{\Odds}(\Rhalf)}\frac{\calN_1\Sigma_1(\Rhalf)}{\calN_0\Sigma_0(\Rhalf)}]^{-1}\ .
\end{equation}
Here $\Rhalf \equiv D \sin{\thetahalf}$ denotes the half-light-radius of the dSph with $\thetahalf$ being the corresponding separation angle, while $\Odds(\Rhalf)$ is the odds of the membership at the half-light-radius. The odds is given by $P(M=1|R=\Rhalf)/P(M=0|R=\Rhalf)$, where $P(M|R)$ is the local membership probability at a specific radius $R$. Here, we note that the parameter $D$ disappear in the photometric likelihood function because $D$ does not depend on $i$. As a result, the likelihood function $\calL_\photo$ has four independent parameters: $\Theta_\photo = \qty{\alpha_0,\,\delta_0,\,\thetahalf,\,\Odds(\Rhalf)}$, where $\alpha_0$ and $\delta_0$ are the right ascension and declination of the centre of the dSph.

\subsubsection{Stellar model selection}
We consider the following two empirical stellar profiles in our analysis; Plummer profile \citep{Plummer1911} and Exponential profile:
\begin{align}\label{eq:stellar_profiles}
\Sigma_1(R) = \begin{cases}
    \displaystyle \frac{1}{\pi \Rhalf^2}{\qty[1+(R/\Rhalf)^2]}^{-2} & (\text{Plummer profile})\\
    \displaystyle \frac{1}{2\pi R_e^2}{\exp(-R/R_e)} & (\text{Exponential profile})\\ 
    \end{cases}  
\end{align}
where $R$ denotes the projected distance from the centre of a dSph, and $R_e$ denotes the exponential radius scale, corresponding to $R_{1/2} = 1.68 R_e$. These two profiles can fit observed stellar profiles of the dSphs very well \citep{Irwin1995, Segall2007, Martin2008}. Those can be analytically de-projected into the three dimensional stellar number density $\nu_1(r)$ as
\begin{equation}
    \nu_1(r) = \begin{cases}
    \displaystyle \frac{3}{4\pi \Rhalf^3} {\qty[1+(r/\Rhalf)^2]}^{-5/2} & (\text{Plummer profile})\\
    \displaystyle \frac{1}{2\pi^2 R_e^3}{K_0(r/R_e)} & (\text{Exponential profile})\\ 
    \end{cases}  
\end{equation}
where $r$ denotes the distance from the centre of a dSph. $K_0(x)$ is the modified Bessel function of the second kind. Note also that the stellar number density is normalised to be $\int\dd{r}\,4\pi r^2\,\nu_1(r) = 1$, so as the projected one, namely $\int\dd{R}\,2\pi R\,\Sigma_1(R) = 1$.

In our analysis, we select the most probable model among the two based on the Bayes factor, namely the ratio of statistical evidences. The statistical evidence of a specific model is defined by the integral of the likelihood function times the prior distribution $\int\dd{\Theta}\calL(\Theta)\pi(\Theta)$, which corresponds to the mean likelihood value of the hypothesis for a given data set. Although the integration of $\int\dd{\Theta}\calL(\Theta)\pi(\Theta)$ is difficult due to large dimensions of $\Theta$, several techniques are developed to evaluate this integration. In this work, we use the Markov Chain Monte Carlo (MCMC) technique to evaluate statistical evidences, as discussed in Section \ref{sec:sampling}.

\subsection{Spectroscopy}\label{sec:spec}
The purpose of this part is to determine DM density profile and $J$-factor of a dSph using the results of the previous part as priors.

\subsubsection{Foreground model selection}
We assume up to three Gaussian components for the Milky Way contamination (the foreground contribution) as already mentioned, namely the thin disk, thick disk, and halo components. In the actual data sample, however, not all of those components are appreciable due to the contribution from a dSph. Hence, in order to determine the number of foreground components in advance, we refer the \Besancon\ model\footnote{\url{https://model.obs-besancon.fr/modele_simuls.php}} \citep{Robin:2003}. We first generate the mock stars according to the model, where the number of generated stars is determined so that it becomes compatible with the actual data. Then, we compare the statistical evidences of N-component foreground models (N=1,2,3) and adopt the most likely one.

\subsubsection{Stellar velocity dispersion and DM density profile}
DM density profiles of dSphs are estimated by comparing observed data of line-of-sight velocity dispersion with theoretical predictions. Assuming that dSphs are spherical and steady systems, dispersion curves of the dSphs are predicted by the following spherical Jeans equation \citep{Binney2008}:
\begin{equation}
    \frac{1}{\nu_1(r)}\pdv{\nu_1(r)\sigma_r^2(r)}{r} + \frac{2\betaani(r)\sigma_r^2(r)}{r} = -\frac{GM(r)}{r^2}\ .
\end{equation}
Here, $G$ is the gravitational constant and $M(r)$ is the enclosed mass within $r$, which is given by $M(r)=\int_0^r\dd{r'} 4\pi {r'}^2 \rho_\text{DM}(r')$ for the case of a dSph due to the fact that the mass is dominated by DM contribution. The velocity dispersion of member stars are defined by $\sigma_r$, $\sigma_\theta$ and $\sigma_\phi$ in general, which denote the dispersion along the radial, polar, and azimuthal directions, respectively.  Now, we can take $\sigma_\theta = \sigma_\phi$ thanks to the spherical symmetry, while the anisotropy parameter $\betaani$ is defined to be $\betaani\equiv 1-(\sigma_\theta^2+\sigma_\phi^2)/(2\sigma_r^2)$.

 We obtain the line-of-sight velocity dispersion $\sigma_\text{l.o.s.}^2(R)$ of dSph member stars by solving the Jeans equation under the assumption of a constant anisotropy parameter $\beta_\text{ani}(r) = \beta_\text{ani}$,
\begin{equation}\label{eq:sigmalos}
    \sigma_\text{l.o.s.}^2(R) = \frac{2}{\Sigma_1(R)}\int_R^\infty\dd{r} \qty(1-\beta_\text{ani}\frac{R^2}{r^2})\frac{\nu_1(r)\sigma_r^2(r)}{\sqrt{1-R^2/r^2}}\ ,
\end{equation}
where the velocity dispersion along the radial direction  $\sigma_r$ is
\begin{equation}\label{eq:sigmar}
    \sigma_r^2(r) = \frac{1}{\nu_1(r)}\int_r^\infty \nu_1(r')\qty(\frac{r'}{r})^{2\beta_\text{ani}}\frac{G M(r')}{{r'}^2} \dd{r'}\ .
\end{equation}

Here, we assume the generalised NFW halo profile \citep{Hernquist:1990,Dehnen1993,Zhao1996} for the DM density profile, 
\begin{equation}
\label{eq:NFW}
    \rho_\text{DM}(r) = \rho_s(r/r_s)^{-\gamma}(1+(r/r_s)^\alpha)^{-(\beta-\gamma)/\alpha},
\end{equation}
where parameters $\rho_s$ and $r_s$ are the scale density and the scale radius of the DM density profile, respectively, while other parameters $\alpha$, $\beta$, and $\gamma$ determine the shape of the profile. For instance, $(\alpha,\,\beta,\,\gamma)=(1,\,3,\,1)$ gives the famous Navarro-Frenk-White (NFW) profile \citep{Navarro:1997}, while $(\alpha,\,\beta,\,\gamma)=(1.5,\,3,\,0)$ (approximately) gives the Burkert profile \citep{Burkert1995}.

\subsubsection{Likelihood function and prior}
In \KI, the likelihood function is defined by the mixture model of dSph member stars and foreground stars, as explicitly shown in Appendix \ref{App:KI17tomodKI17}, which is based on the simultaneous probability density $P(v,R)$ with $v$ and $R$ being the stellar line-of-sight velocity and the radius from the centre of the dSph, respectively. The likelihood function is indeed proved to successfully reproduce original input parameters by analysis using mock data, as was shown in \KI.

To analyse actual observed data, however, we should consider the spatial sampling bias of observed stars \citep{Martinez2011}. We therefore use a improved likelihood function for spectroscopic samples based on the conditional probability $P(v|R)=P(v,R)/\int\dd{R}P(v,R)$ rather than $P(v,R)$ itself,
\begin{multline}
\calL_\spec(\Theta_\spec) = 
    \prod_i\biggl[ s(R_i) \mathcal{G}\qty[v_i; v_1,\sqrt{\sigma_1^2(R_i)+\delta\sigma_i^2}] \\ 
    + [1 - s(R_i)] \sum_c \pi_c \mathcal{G}\qty[v_i; v_{0,c},\sqrt{\sigma_{0,c}^2+\delta\sigma_i^2}] \biggr],
\end{multline}
where $\calG[v;\mu,\sigma]$ is the Gaussian function whose mean and standard deviation are given by $\mu$ and $\sigma$, respectively. Here, $\delta\sigma_i$ is the observational error of the $i$-th star.\footnote{In \KI, we ignored this observational error because it does not cause significant difference in $J$-factor estimation. However, we find that this term improves the performance of the MCMC because the singularity of the likelihood function at $\sigma_{0,c}\to 0$ can be removed by introducing $\delta\sigma_i$.} The mean velocity of member stars are denoted by $v_1$, while the velocity dispersion of the stars is $\sigma_1(R)$, which  is nothing but $\sigma_\text{l.o.s.}(R)$ defined in equation (\ref{eq:sigmalos}). On the other hand, $v_{0,c}$, $\sigma_{0,c}$ and $\pi_c$ denote the mean velocity, velocity dispersion and weight of the $c$-th foreground component, respectively, where the coefficients $\pi_c$ are normalised to be $\sum_c \pi_c = 1$. At last, $s_R$ denotes the local membership probability of stars at the radius $R$ (see also Appendix \ref{App:KI17tomodKI17}), which is given as follows:
\begin{equation}
    s(R) = \qty[1 + \frac{1}{\Odds(\Rhalf)}\frac{\Sigma_1(\Rhalf)/\Sigma_1(R)}{\Sigma_0(\Rhalf)/\Sigma_0(R)}]^{-1}\ .
\end{equation}
Parameters of the spectroscopic likelihood function are $\Theta_\spec = \Theta_\photo \cup \qty{r_s,\,\rho_s,\,\alpha,\,\beta,\,\gamma,\,\beta_\text{ani},D,v_1} \cup_c \qty{\pi_c,\,v_{0,c},\sigma_{0,c}}$. Here, we should note that the number of independent parameters are $|\Theta_\spec| - 1$ due to $\sum_c \pi_c=1$; e.g. we have $4+8+3\times3-1=20$ independent parameters for the three-component foreground model.

We introduce Gaussian, flat, log-flat priors for these parameters: We consider Gaussian priors, $\pi(\Theta_\photo) = \calG[\Theta;\mu_\Theta,\sigma_\Theta]$ and $\pi(D) = \calG[D;\mu_D,\sigma_D]$, for $\Theta_\photo$ and $D$, respectively. Here $\mu_\Theta$ and $\sigma_\Theta$ is the median and the half of the 68\% quantile of the posterior distribution obtained by the statistical analysis in the photometry part mentioned in the previous subsection, while $\mu_D$ and $\sigma_D$ are the observed distance and its error of a dSph taken from \citet{McConnachie2012a}. For the DM halo parameters and the anisotropy parameter, we use the flat and log-flat priors over following ranges:
\begin{gather*}
    -4 \leq -\log_{10}(\rho_s/[\Msun\text{pc}^{-3}]) \leq 4\ , \\
    0 \leq -\log_{10}(r_s/[\text{kpc}]) \leq 5\ , \\
    0.5 \leq \alpha \leq 3\ , \\
    3 \leq \beta \leq 10\ , \\
    0 \leq \gamma \leq 1.2\ , \\
    -1 \leq -\log_{10}(1-\betaani) < 1\ ,
\end{gather*}
which are the same as those adopted in \cite{Geringer-Sameth2015}. We impose the flat prior for $v_1$ over the following range:
\begin{equation*}
-10^3 < v_1/[\text{km s}^{-1}] < 10^3\ .
\end{equation*}
For the foreground spectroscopic parameters ($\pi_c$, $v_{0,c}$ and $\sigma_{0,c}$), we impose flat priors over the following ranges:
\begin{gather*}
    0 \leq \pi_c \leq 1 \, {\rm with} \, \pi_1>\pi_2>\pi_3\ (\text{see also Section \ref{sec:sampling}}), \\
    -10^4 \leq v_{0,c}/[\text{km s}^{-1}] \leq 10^4\ , \\
    0 \leq \sigma_{0,c}/[\text{km s}^{-1}] \leq 10^4\ .
\end{gather*}

\subsection{Sampling algorithm}\label{sec:sampling}
Our likelihood functions and posteriors have many parameters. In particular, the spectroscopic one has more than ten parameters. The Markov Chain Monte Carlo (MCMC) method is known to enable us to generate parameter samples whose distribution satisfies such a multidimensional function. For example, the Metropolis-Hastings algorithm \citep{Metropolis1953, Hastings:1970aa} is known as a simple MCMC algorithm. It requires, however, the tuning of hyperparameters, such as the step width and correlation matrix of the random walk in the parameter space. In our study, we use more sophisticated MCMC sampler, the Affine Invariant Ensemble Sampler implemented by \texttt{emcee} \citep{Foreman-Mackey2013}, which provides us easy interfaces to make a MCMC code in {Python} without any hyperparameter tuning except for the step number and the number of \textit{walkers} (parallelised MCMC sampler). We perform parameter samplings with $\order{10^6}$ steps using this sampler.

It is important to point out here that the spectroscopic likelihood function has a permutation symmetry which exchanges the foreground components with their parameters (mean, dispersion, and weight). Such a symmetry is known to cause the label-switching problem \citep{Jasra2005MarkovModeling} by the multi-modality of the likelihood function corresponding to the symmetry. Although an additional ordering condition (e.g. $\pi_1 > \pi_2 > \pi_3$, $\mu_1 > \mu_2 > \mu_3$ or $\sigma_1 > \sigma_2 > \sigma_3$) can break the symmetry, these procedures yield accidental local maxima as by-products due to the hard cut of the parameter space. These local maxima trap a part of the MCMC samplers in low-likelihood regions and distort the shape of the posterior distribution functions. To resolve the problem, we impose the weight ordering condition ($\pi_1 > \pi_2 > \pi_3$) as denoted above and remove MCMC samples trapped around local maxima having significantly small (by the factor of $<10^{-5}$) posterior values.

To evaluate the statistical evidence of each model, we adopt the widely applicable Bayesian information criterion (WBIC) as an approximation of the (minus-log) evidence. The WBIC can be easily computed by MCMC samples and it is valid even for singular model such as Gaussian mixture model and our spectroscopic likelihood function (see Appendix \ref{sec:wbic} for more details). To evaluate the WBIC, we perform MCMC samplings with $\order{10^6}$ steps.

\section{Data}
\begin{figure*}
    \centering
    \myminipage{0.3\hsize}{\includegraphics[width=\hsize]{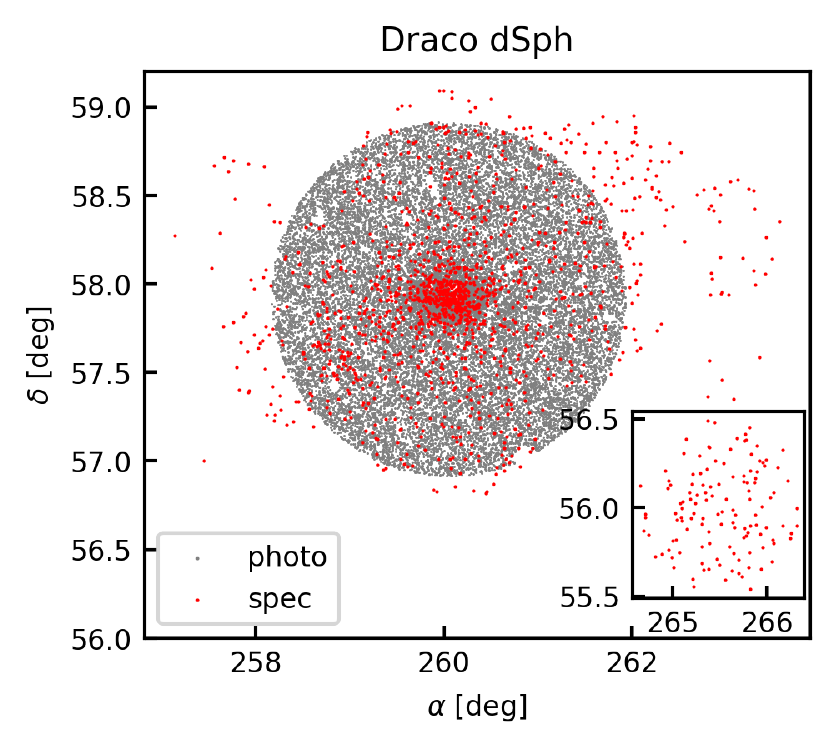}}
    \myminipage{0.3\hsize}{\includegraphics[width=0.9\hsize]{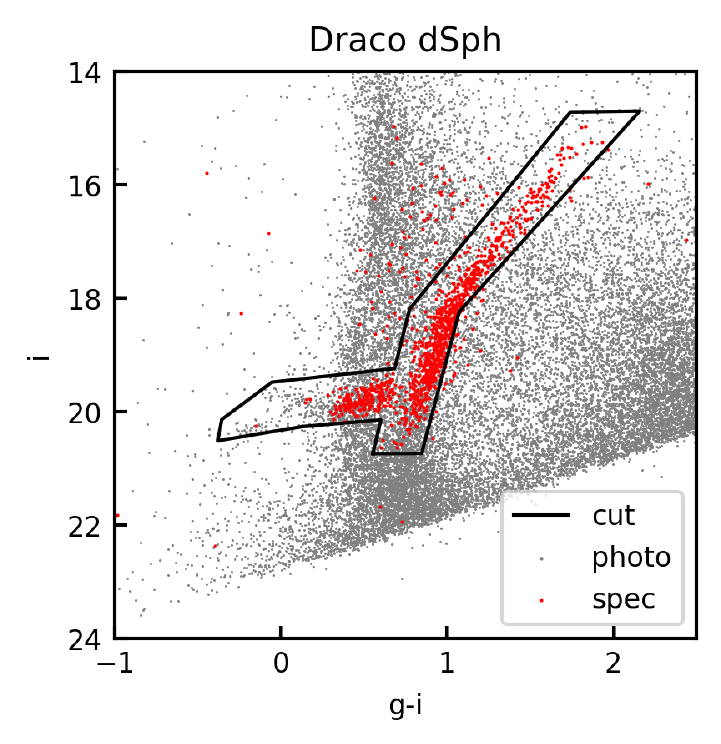}}
    \myminipage{0.3\hsize}{\includegraphics[width=0.9\hsize]{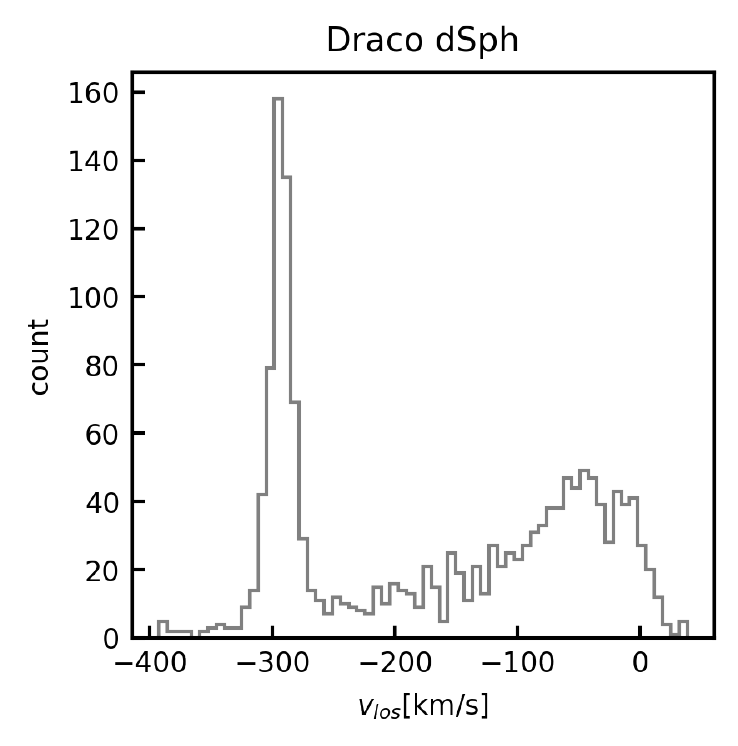}}
    \myminipage{0.3\hsize}{\includegraphics[width=\hsize]{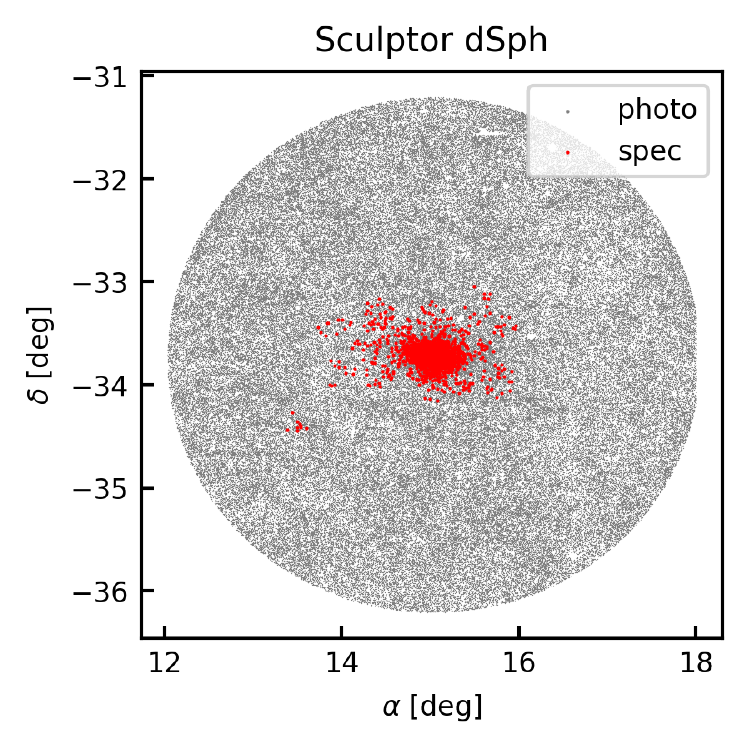}}
    \myminipage{0.3\hsize}{\includegraphics[width=0.9\hsize]{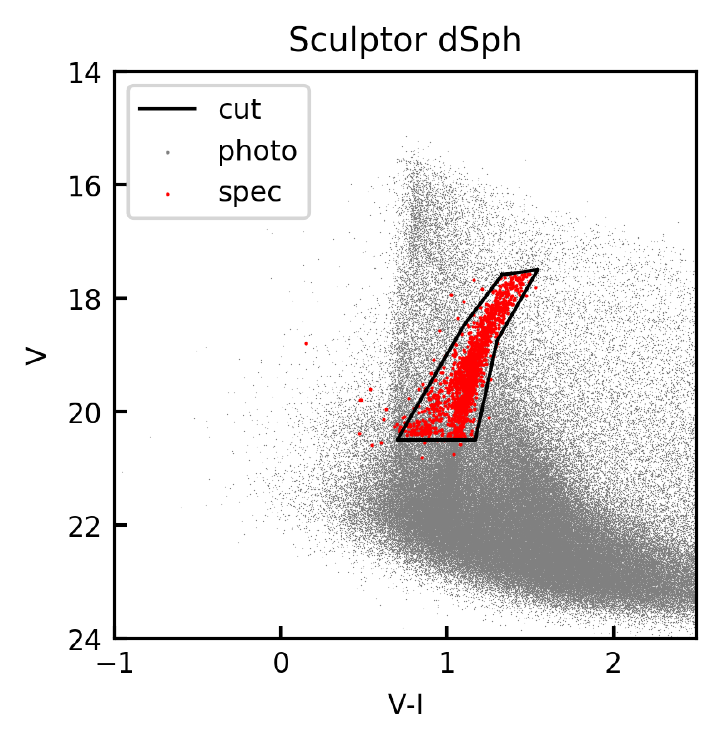}}
    \myminipage{0.3\hsize}{\includegraphics[width=0.9\hsize]{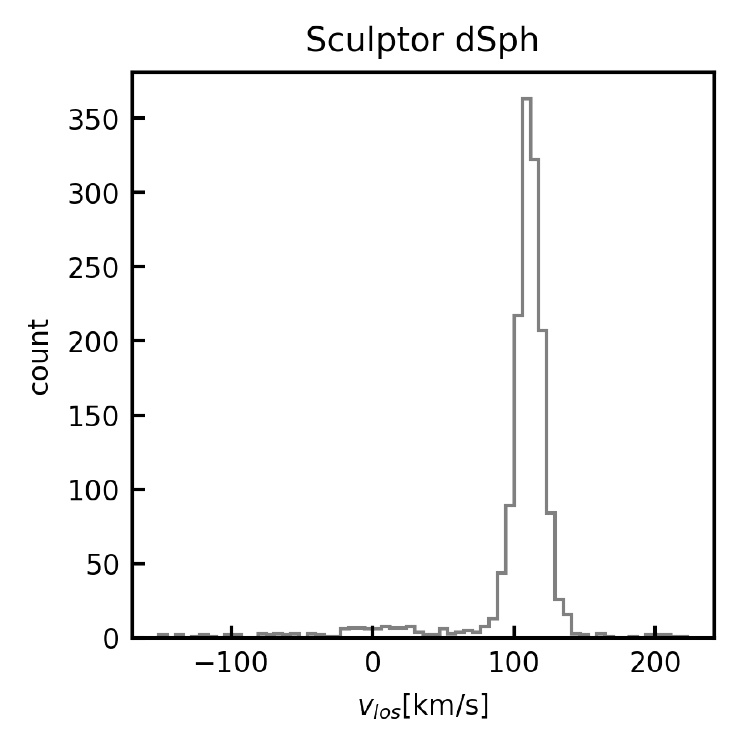}}
    \myminipage{0.3\hsize}{\includegraphics[width=\hsize]{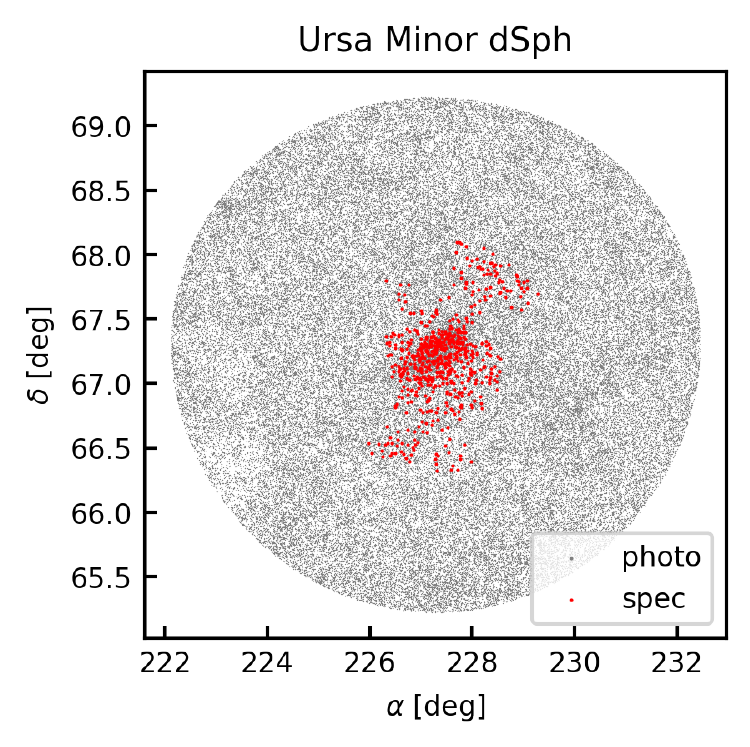}}
    \myminipage{0.3\hsize}{\includegraphics[width=0.9\hsize]{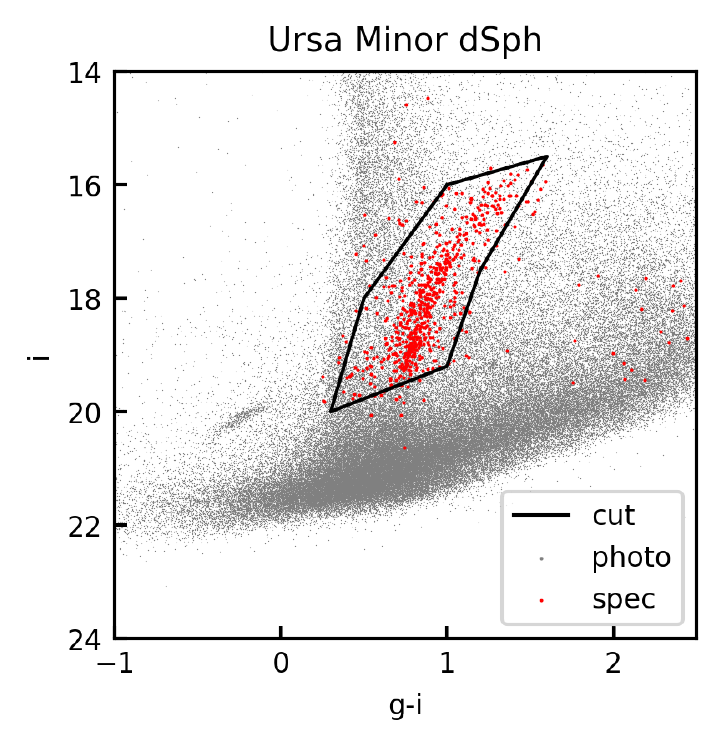}}
    \myminipage{0.3\hsize}{\includegraphics[width=0.9\hsize]{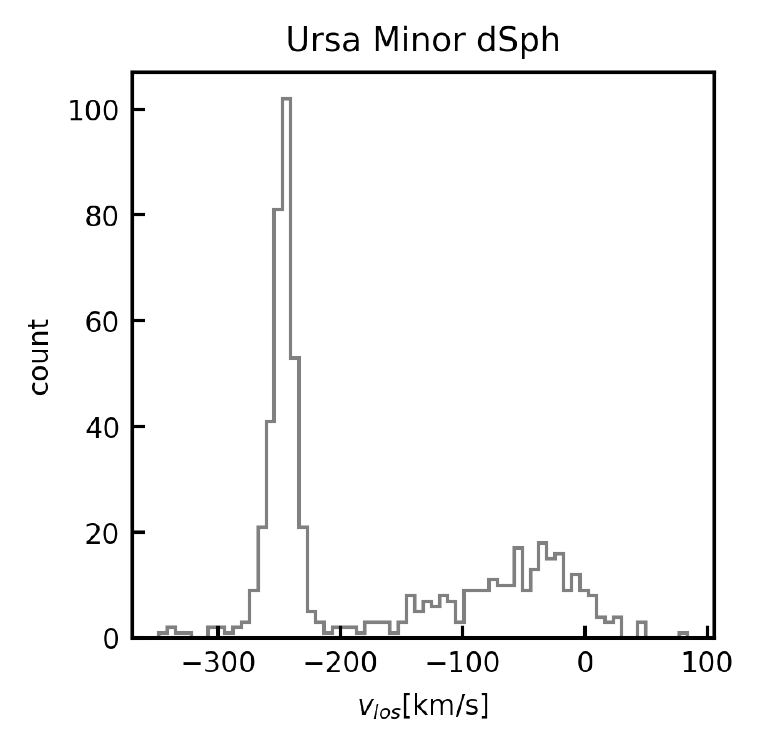}}
    \caption{\small Photometric and spectroscopic samples of Draco (\textbf{top row}), Sculptor (\textbf{middle row}) and Ursa Minor (\textbf{bottom row}). Grey and red dots on scatter plots correspond to the photometric and spectroscopic samples, respectively. \textbf{Left column}: Spacial distribution of stars in the equatorial coordinate system. \textbf{Centre column}: Colour-magnitude diagram (CMD) with corrections of the reddening. The magnitudes of the spectroscopic samples are obtained from the photometric samples through the catalogue matching. We use the stars in the black polygons for the CMD cut. Note that we use different colour-magnitude systems among the dSphs (SDSS $g$- and $i$-band for Draco, Johnson $V$- and $I$-band for Sculptor, Pan-STARRS $g$- and $i$-band for Ursa Minor, respectively). \textbf{Right column}: The histogram of the velocity distribution of the spectroscopic samples we have before the CMD cut (red points in left and centre columns).}
    \label{fig:CMDcut}
\end{figure*}

In this section, we discuss the sources of data sets used in our analysis for photometric and spectroscopic samples. We also explain the pre-processing of these data sets, which is also shown in Fig.\,\ref{fig:CMDcut}. 

\subsection{Data sources}\label{sec:DataSource}
The sources of photometric and spectroscopic samples for each dSph are summarised in Table\,\ref{DataSource}. \textbf{Photometry}: For Draco, Sculptor and Ursa Minor, we have referred to SDSS DR14 \citep{Abolfathi2018}, DES DR1 \citep{Abbott2018} and Pan-STARRS DR1 \citep{Chambers2016}, respectively, to obtain the position (right ascension, declination) and the magnitudes ($g$, $r$, $i$, $z$ and $y$ bands) of stars. \textbf{Spectroscopy}: For Draco, we have used stellar-kinematic samples of MMT/Hectochelle observation \citep{Walker2015a}, which provide positions and line-of-sight velocities of the stars. For Sculptor, we have referred to the result of Magellan/MMFS survey \citep{Walker2009d} to obtain the positions, line-of-sight velocities and $V-$ and $I-$band magnitudes of the stars. For Ursa Minor, we have obtained the position and velocity data from the observation of MMT/Hectochelle telescope \citep{Spencer2018}.

\begin{table*}
	\centering
	\label{tab:DataSource}
	\begin{tabular}{llrlr} % four columns, alignment for each
		\hline
		dSph & Photometry & $R_\photo$ [$\deg$] & Spectroscopy & $r_\text{max}$ [pc]\\
		\hline
		Draco & SDSS DR14 \citep{Abolfathi2018} & 1.0 &  MMT/Hectochelle \citep{Walker2015a} & 1866 \\
		Sculptor & DES DR1 \citep{Abbott2018} & 2.5 & Magellan/MMFS \citep{Walker2009d} & 2673\\
		Ursa Minor & Pan-STARRS DR1 \citep{Chambers2016} & 2.0 & MMT/Hectochelle \citep{Spencer2018} & 1580\\
		\hline
	\end{tabular}
    \caption{\small Photometric and spectroscopic data, where $r_\text{max}$ is the radius of the outermost star \citet{Geringer-Sameth2015}. See Section \ref{sec:DataSource} for more details.}
    \label{DataSource}
\end{table*}

We have used samples (stars) within the radius $R<R_\text{photo}$, where $R_\text{photo}=1.0 \deg$, $2.5 \deg$ and $2.0 \deg$ for Draco, Sculptor and Ursa Minor, respectively. We have decided these radii to include both the outer-most likely member stars of each dSph and the foreground stars that are sufficient to carry out the multi-component analysis described in the previous section. Note also that the radii are specifically optimised to ensure that the spatial and velocity distribution of the foreground stars can be assumed to be uniform.

\subsection{Cross-matching among photometric \& spectroscopic data}\label{sec:crossmatching}
To determine the magnitudes of spectroscopic stars, we looked for the closest photometric star on the equatorial coordinates for each spectroscopic star and regard the two stars identical. If the spectroscopic star is located much away from the nearest photometric star (farther than $5''$), it is removed in the spectroscopic analyses.\footnote{Almost all of the spectroscopic star samples can be matched to the photometric samples. Indeed, for each dSph, only a few per-cent of the samples are removed from data used in subsequent spectroscopic analyses. }

We note that our spectroscopic samples of Sculptor have the information of colour-magnitudes measured in the Johnson-Morgan system. In order to cross-check our identification, we have used conversion formula of the DES DR1\footnote{\url{https://des.ncsa.illinois.edu/releases/dr1/dr1-faq}} between the DES system and the SDSS system and that of \citet{Abbott2018,Drlica-Wagner2017a} between the SDSS system and the Jonson-Morgan system. Then, we have confirmed the consistency between the original magnitudes of the spectroscopic samples and converted magnitudes of the corresponding photometric samples.

\subsection{Colour-magnitude cut}\label{sec:cmdcut}
When we estimate membership probabilities of different sample sets such as photometric and spectroscopic ones, we should keep in mind that the membership probability generally depends on the choice of colour-magnitude cuts because the colour-magnitudes of member stars are different from those of foreground stars. In order to guarantee the population of stars to be equal between the photometric and the spectroscopic samples, we impose the same cuts on the colour-magnitude diagram (CMD) in our analysis as shown in Fig.\,\ref{fig:CMDcut}. The CMD cut for the Draco dSph is based on that of \citet{Walker2015a}, which includes red giants and horizontal branch stars. For the Sculptor and Ursa Minor dSphs, we define our cuts with simple polygons including most of the spectroscopic samples.

\section{Results}
\label{sec:Results}

\begin{figure*}
    \centering
    \myminipage{0.3\hsize}{\includegraphics[width=\hsize]{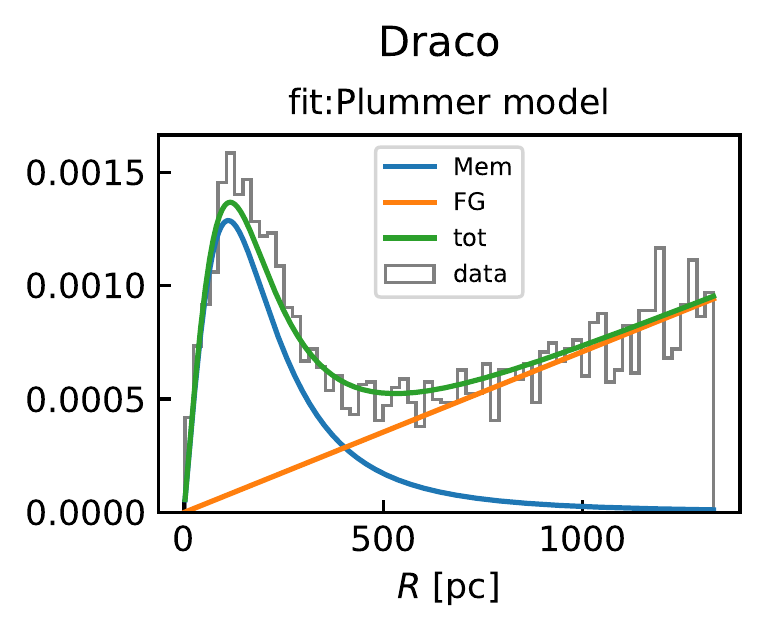}}
    \myminipage{0.3\hsize}{\includegraphics[width=\hsize]{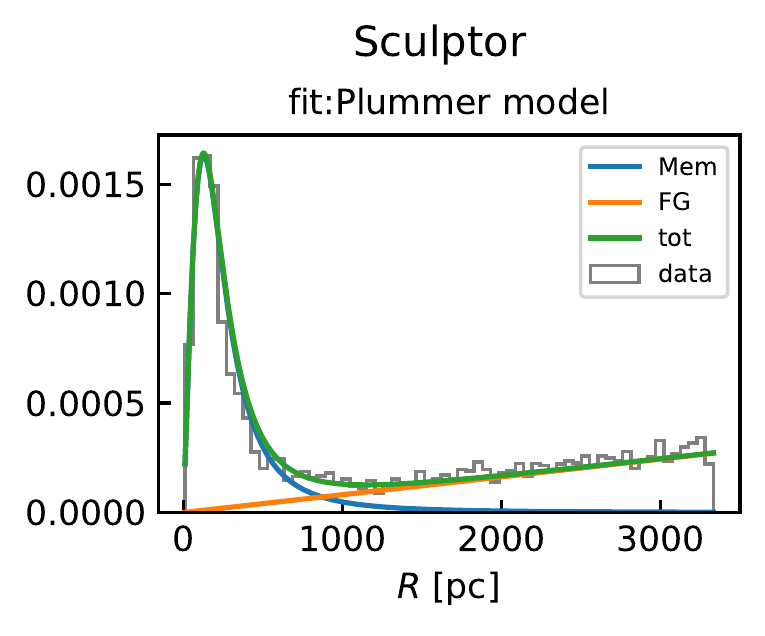}}
    \myminipage{0.3\hsize}{\includegraphics[width=\hsize]{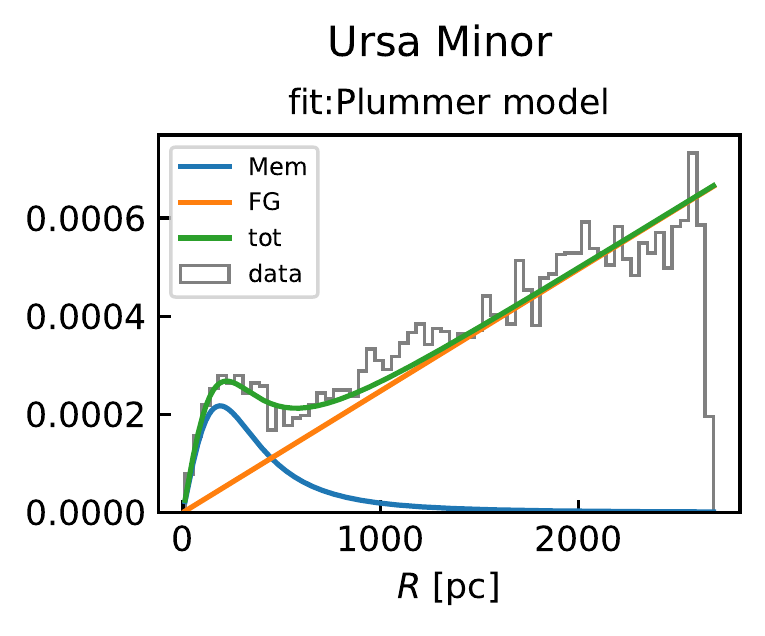}}
    \caption{\small Result of fitting for Draco (\textbf{left}), Sculptor (\textbf{centre}) and Ursa Minor (\textbf{right}) dSphs in the photometric analysis. The grey histograms in the panels show the normalised surface density profile of photometric stars. On the other hand, the coloured lines in the panels show the number density $\Sigma(R)$ times $2\pi R$ of member stars (blue), foreground stars (orange) and all stars (green), respectively, based on the maximum a posteriori (MAP) parameters, which are identical to those obtained by maximum likelihood estimation because we are using flat priors for the photometric parameters $\Theta_\photo$ in our analysis.}
    \label{fig:photofit}
\end{figure*}

As the result of the stellar density profile selection, the Plummer model is accepted for all the dSphs, though the Bayes factors do not vary much as $\ln\text{BF}\simeq2\sim6$ among the dSphs. It is noteworthy that this selection process will be more important when we consider a stellar model out of more complicated ones because our procedure gives a systematic approach to choose the best one. To confirm the validity of our photometry analysis, we show in Fig.\,\ref{fig:photofit}  the comparison between the observed surface density and the prediction of the adopted stellar model (obtained by the posterior probability density of the photometry analysis) for each dSph. The grey histograms in the panels of the figure show the binned surface densities of the photometric samples integrated over a ring with a radius $R$, while coloured lines shows those obtained by the result of our analysis. Thanks to a sufficient number of the photometric samples of ${\cal O}(10^4)$, their PDFs converge into Gaussian-like distributions without any prior dependence, which indicates that the result of our photometric analysis with Bayesian statistics is also expected to be achieved even by the analysis with the frequentist statistics. The lines shown in the figure are from estimated number densities computed at the maximum a posteriori parameters, which is nothing but the counterpart of the maximum likelihood estimation or chi-square fittings of the samples in the frequentist statistics.

In the spectroscopic analysis, the WBIC test for the foreground model selection shows that only two foreground components based on the \Besancon\ model are sufficient to fit foreground star distributions of the dSph candidates (Draco, Ursa Minor and Sculptor). This is partially because the foreground stars belonging to the thin disk component have much different colour-magnitude properties from those of the member stars, and thus our cut on the colour-magnitude diagram can remove the thin disk stars. Hence, we use the two components model in the spectroscopic analysis.\footnote{It is worth notifying that the foreground model selection has the CMD cut dependence as well as the membership probability, because each foreground component has its typical magnitudes and velocity distribution. The model selection procedure therefore should be repeated each time in future analyses whenever we use a different sample set with a different CMD cut.}

\begin{figure*}
    \centering
    \includegraphics[width=\hsize]{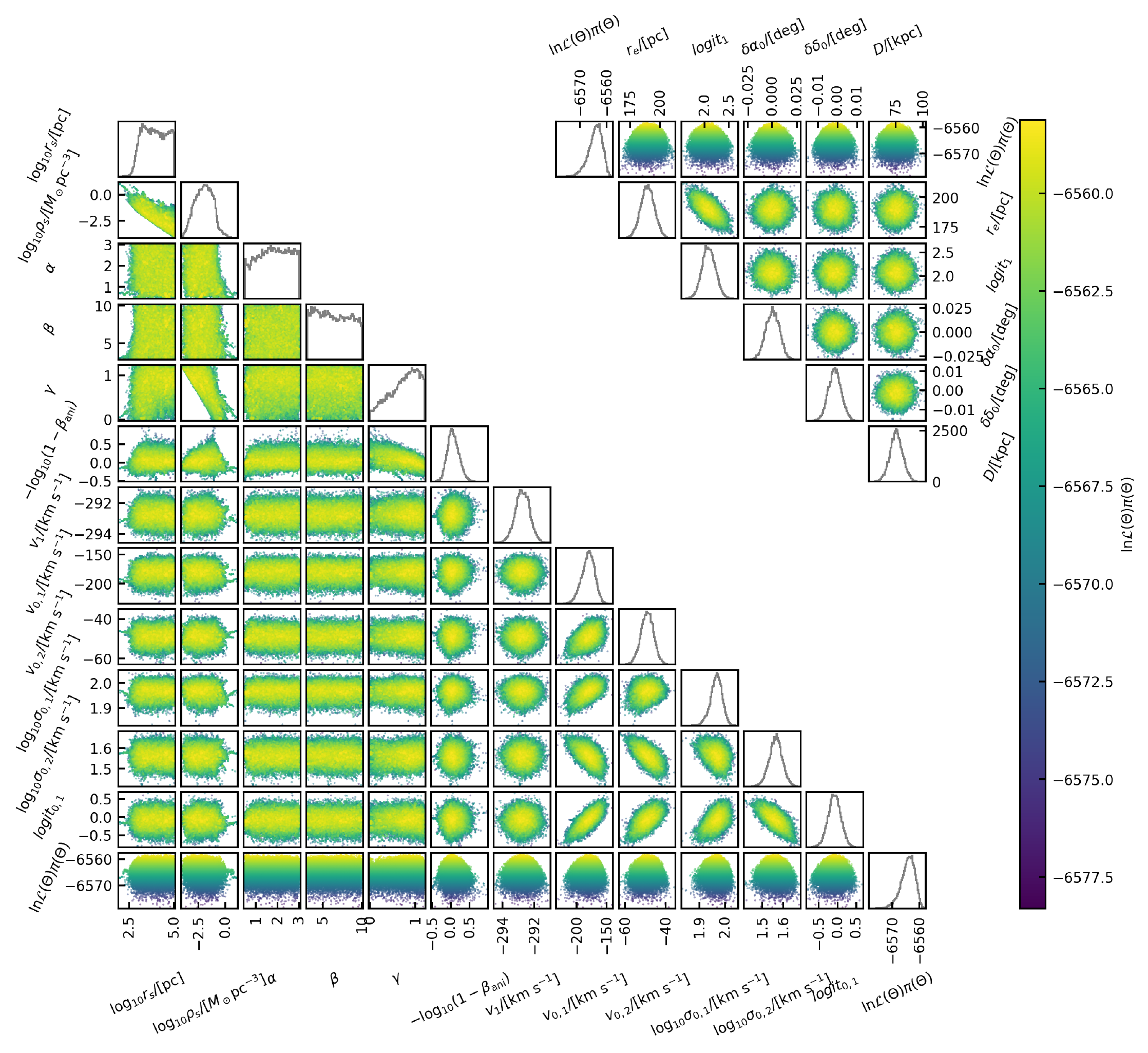}
    \caption{\small Posterior probability density and correlation matrix for Draco parameters. The upper right triangle corresponds to photometric parameters $\Theta_\photo$ as well as a spectroscopic parameter $D$, while the lower left triangle corresponds to the spectroscopic ones $\Theta_\spec$ except $D$. For illustration, we convert the velocity dispersion parameters $\sigma_i$ into $\log_{10}\sigma_i$, odds parameter $\Odds(\Rhalf)$ into $\mathrm{logit}_1 \equiv \ln(\Odds(\Rhalf))$ and ${\pi_i}$ into $\mathrm{logit}_{0,i} \equiv \ln({\pi_i}/(1-{\pi_i}))$.}
    \label{fig:scattermatrix}
\end{figure*}

\begin{figure*}
    \centering
    \myminipage{0.3\hsize}{\includegraphics[width=\hsize]{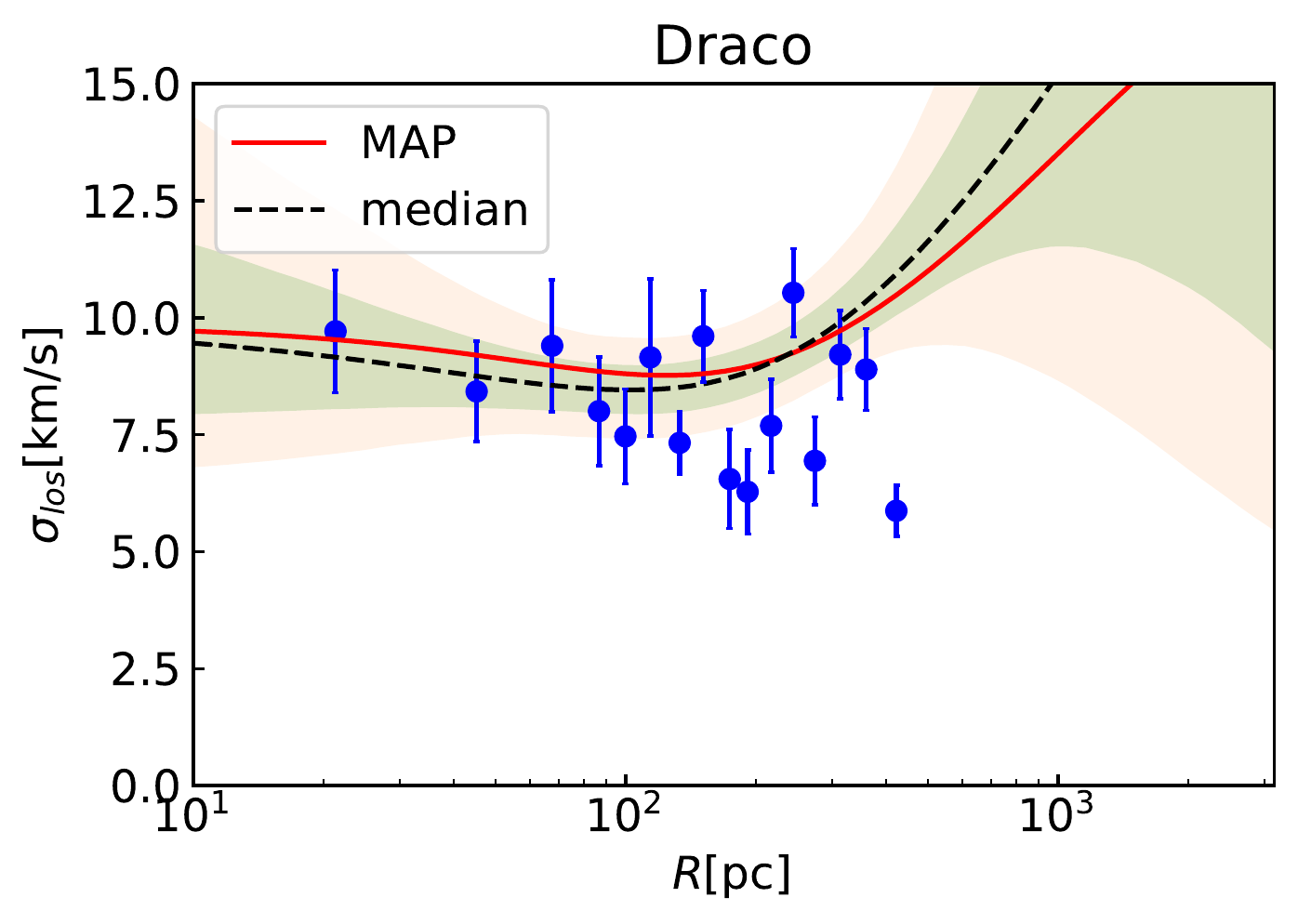}}
    \myminipage{0.3\hsize}{\includegraphics[width=\hsize]{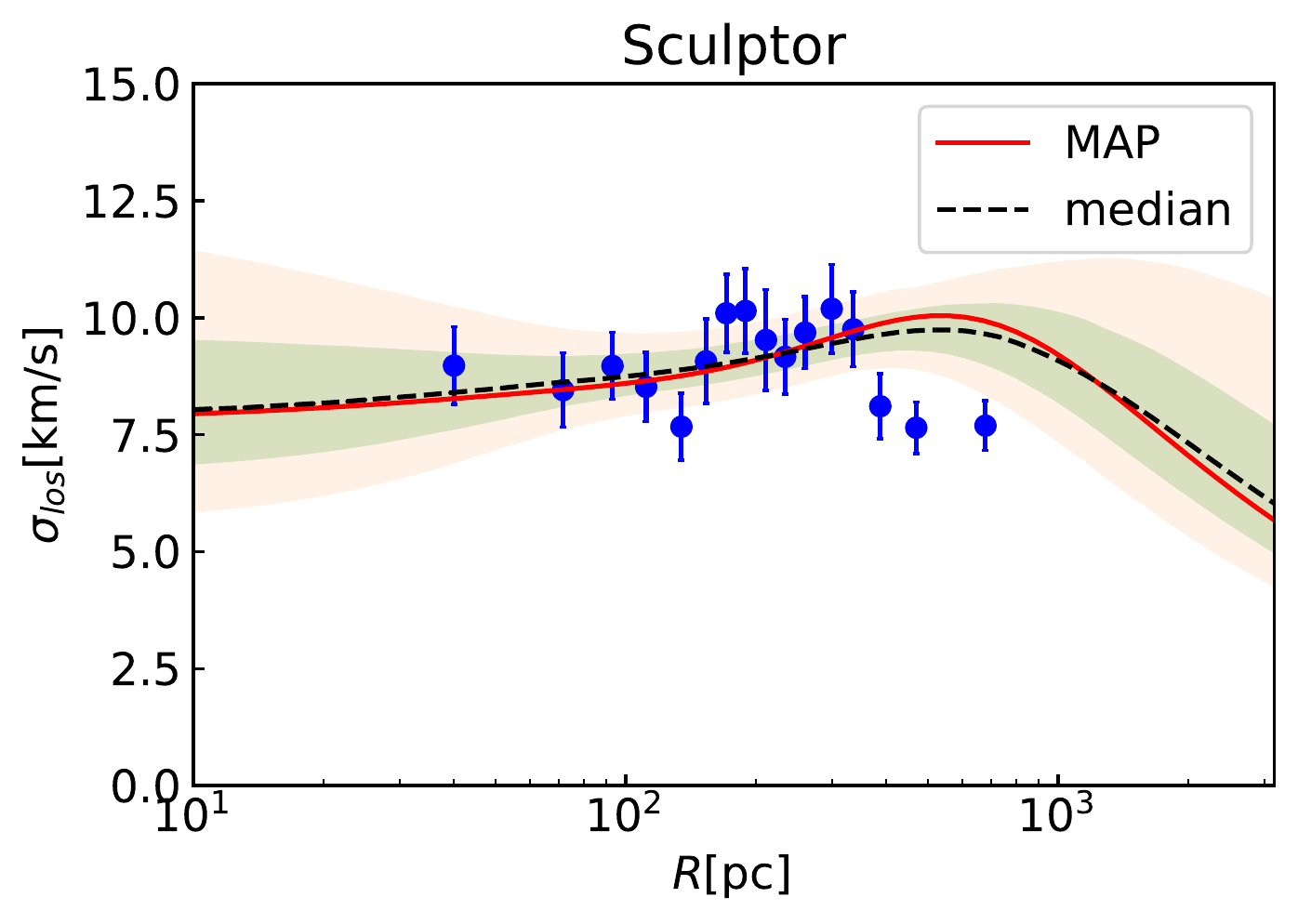}}
    \myminipage{0.3\hsize}{\includegraphics[width=\hsize]{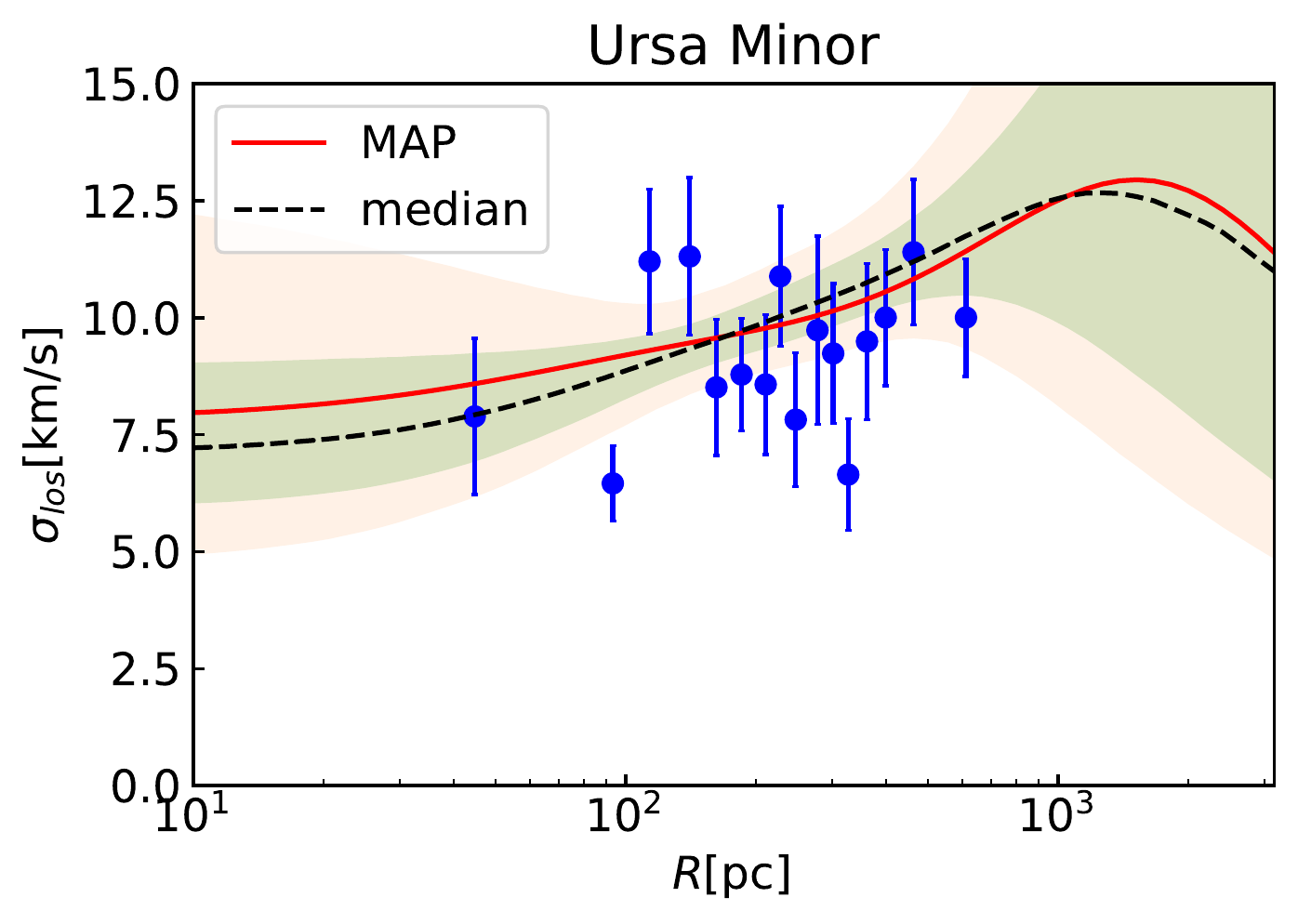}}
    \caption{\small Comparison between observed and estimated velocity dispersion of the member stars for Draco (\textbf{left}), Sculptor (\textbf{centre}) and Ursa Minor (\textbf{right}). Blue points are observed dispersion with error bars obtained by a bootstrap sampling. Dashed lines denote the median values of the estimated dispersion associated with Bayesian credible intervals of 68\% and 95\% shown by green and yellow bands. Maximum a posteriori (MAP) lines are also shown by red lines.}
    \label{fig:specfit}
\end{figure*}

Posterior probability density functions (PDFs) and their correlations are shown in \cref{fig:scattermatrix,fig:scattermatrix_scl,fig:scattermatrix_umi} for Draco, Sculptor and Ursa Minor, respectively. In these figures, for illustration, we convert the velocity dispersion parameters $\sigma_i$ into $\log_{10}\sigma_i$, odds parameter $\Odds(\Rhalf)$ into $\mathrm{logit}_1 \equiv \ln[\Odds(\Rhalf)]$ and $\pi_i$ into $\mathrm{logit}_{0,i} \equiv \ln[{\pi_i}/(1-{\pi_i})]$. On the other hand, Fig.\,\ref{fig:specfit} shows the comparison between observed and estimated velocity dispersion of the member stars, obtained by the PDF of the spectroscopic analysis. Here, blue points with error bars are the observed velocity dispersion calculated by binned samples with the median membership probability $\langle P_M\rangle \geq 0.5$ (member-like stars), while corresponding error bars are obtained by a bootstrap sampling of the stars in the bins. Dashed lines denote the median values of the estimated velocity dispersion, while green and yellow bands are Bayesian credible intervals of 68\% and 95\%, respectively.  We also show maximum a posteriori (MAP) lines by red line as an analogy of the maximum likelihood estimation (MLE) or simple chi-square fitting.

\begin{figure}
    \centering
    \includegraphics[width=0.8\hsize]{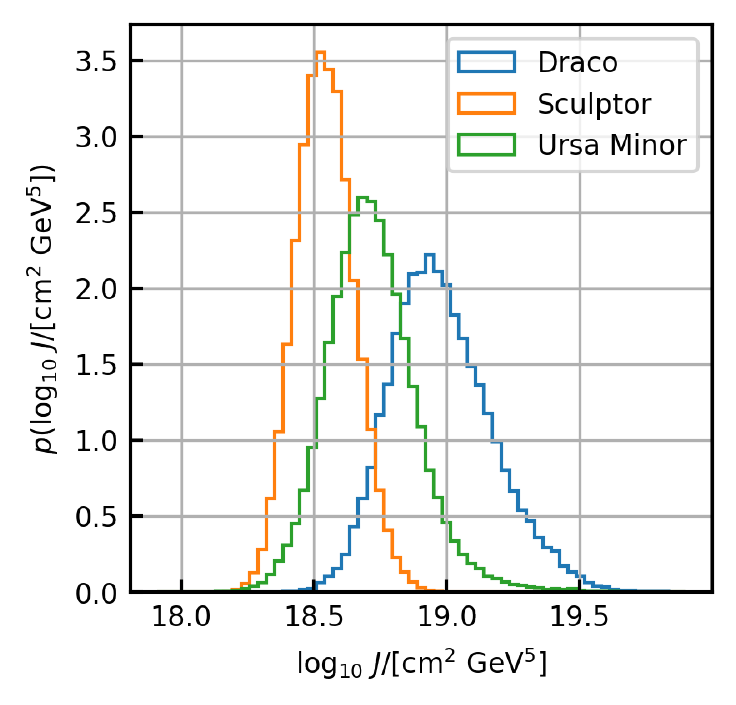}
    \caption{The posterior density function (PDF) of the $J$-factor for each dSph, obtained by the MCMC calculation. Blue, orange and green lines correspond to the PDFs for Draco, Sculptor and Ursa Minor, respectively. Here the vertical axis is normalised to satisfy $\int\dd{(\log_{10}J)} p(\log_{10}J) = 1$.}
    \label{fig:Jhist}
\end{figure}

We show our result of $J$-factor estimation in Table\,\ref{tab:Jresults} and Fig.\,\ref{fig:Jhist} as numbers and posterior probability density functions, respectively. Here, the value of the $J$-factor for each dSph is  obtained by integrating the factor within $0.5 \deg$ solid angle, which is nothing but the standard choice of the $J$-factor estimation \citep{Ackermann2015a}. Note also that we take the radius of the outermost star shown in Table\,\ref{tab:DataSource}, which is given by \citet{Geringer-Sameth2015} as a truncation radius of the $J$-factor estimation to make the estimation the most conservative for the indirect dark matter detection.\footnote{See Fig.\,\ref{fig:truncation} for the impact of the truncation choice on $J$-factor values.} 

\begin{table}
    \centering
    {\renewcommand{\arraystretch}{1.5}
    \begin{tabular}{lcc}
        \hline
        dSph & $\nu,\,\Sigma$ & $\log_{10} (J(0.5^\circ)/ [\text{GeV}^2\text{cm}^{-5}])$ \\ 
        \hline
        Draco & Plummer & $18.96_{-0.17}^{+0.21}$ \\
        Sculptor & Plummer & $18.53_{-0.11}^{+0.12}$ \\
        Ursa Minor & Plummer & $18.75_{-0.13}^{+0.17}$ \\
        \hline
    \end{tabular}
    \renewcommand{\arraystretch}{1.0}}
        \caption{\small Result of $J$-factor estimation. The median value of the posterior probability density function is shown for each dSph. Lower and upper errors correspond to the $1\sigma$ range of the PDF (16th and 84th percentiles).}
    \label{tab:Jresults}
\end{table}

\begin{figure*}
    \centering
    \includegraphics[width=0.8\hsize]{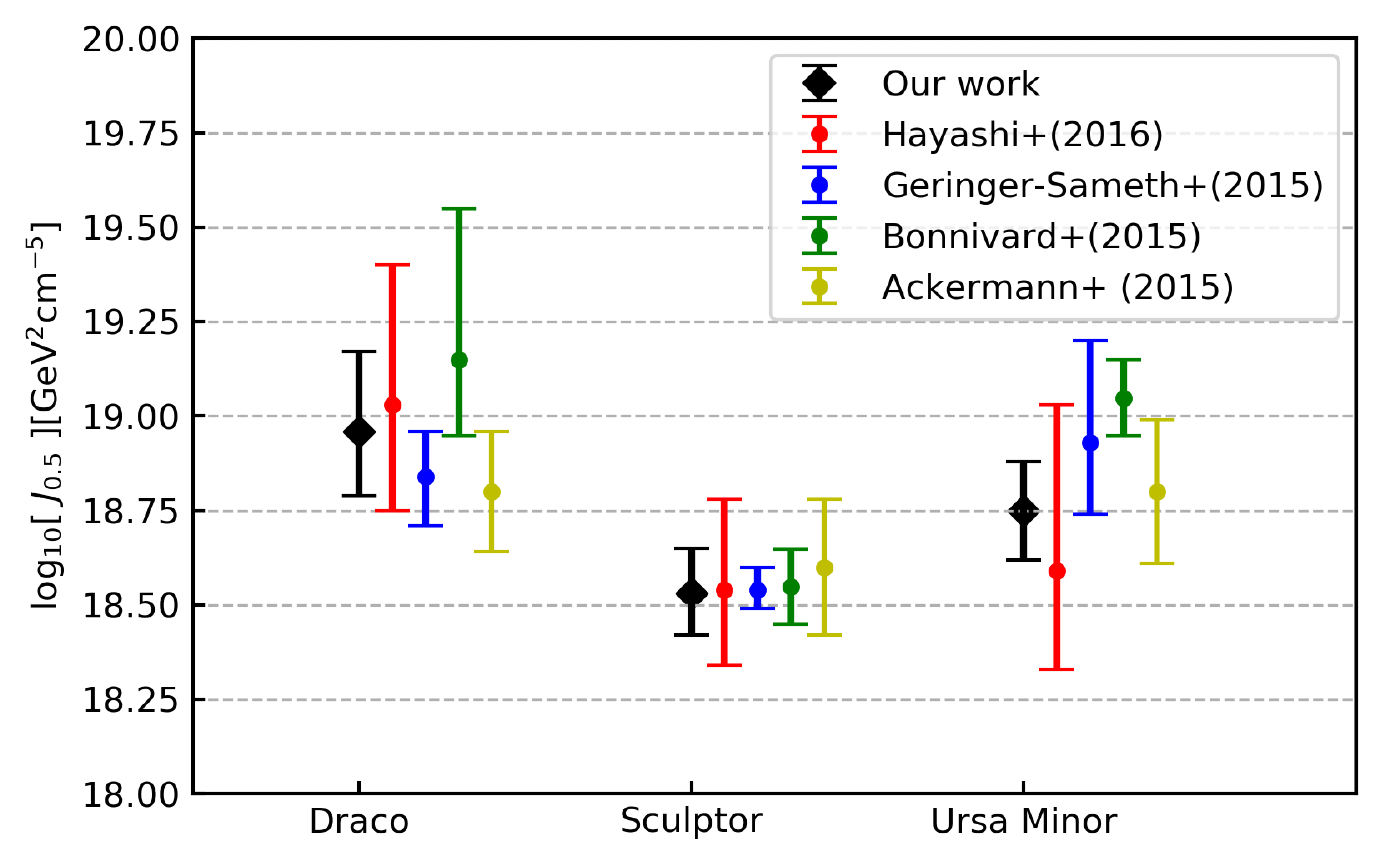}
    \caption{\small Comparison among various $J$-factor estimations. Our result is shown by the black dots error bars. We also show results estimated by other studies \citep{Geringer-Sameth2015,Bonnivard2015a,Ackermann2015a} with blue, green and yellow ones, respectively. For the red ones, we present our (previous) result obtained by performing a more generalised fitting (adopting the generalised NFW profile) than the original one in \citet{Hayashi2016}.}
    \label{fig:Jresults}
\end{figure*}

We compare our result to those of other studies in Fig.\,\ref{fig:Jresults} adopting different methods to treat the contamination effect. Dots and error bars in the figure denote median values and  $68$ percent (1$\sigma$) Bayesian credible intervals of the posterior probability densities. Our result is shown by the black bars. The red, green and blue bars show results of \citet{Hayashi2016,Bonnivard2015a,Geringer-Sameth2015}, respectively, where they adopted the membership selection based on the EM algorithm to remove contaminated stars: \citet{Hayashi2016} considered axisymmetric stellar and DM distributions assuming the generalised NFW profile. In \citet{Bonnivard2015a}, generalised stellar and DM density distributions are assumed with partially taking the uncertainties of dSph triaxiality into account. \citet{Geringer-Sameth2015} performed similar analysis to ours except the treatment of the contamination effect. The yellow bars are from \citet{Ackermann2015a}, where the NFW profile is assumed. Authors also assumed a linear relationship among the total luminosity, maximum circular velocity and radius corresponding to the maximum circular velocity, as inferred by \citet{Martinez:2013els}. We summarise these five studies in Table\,\ref{tab:works}.

\begin{table*}
    \centering
    \begin{tabular}{llllll}
    \hline
    Work                & Symmetry      & $\nu_1$, $\Sigma_1$ & $\rho_\text{DM}$ & $\betaani$ & FG contami. \\
    \hline
    Our study            & Spherical     & Plummer/exp.  & Generalised NFW   & Constant     & Mixture model\\
    \cite{Hayashi2016}  & Axisymmetric  & Plummer       & Generalised NFW   & Constant     & $P_M$ cut \\
    \cite{Geringer-Sameth2015} & Spherical & Plummer   & Generalised NFW   & Constant     & $P_M$ cut \\
    \cite{Bonnivard2015a} & Spherical (triaxial) & Generalised NFW & Generalised Einasto & \citet{Baes:2007tx} & $P_M$ cut \\
    \cite{Ackermann2015a} & Spherical & - & NFW & - & Hierarchical modelling \\ 
    \hline
    \end{tabular}
    \caption{\small Comparison of analysis methods in Fig.\,\ref{fig:Jresults}. Note that the stellar and anisotropy models of \citet{Ackermann2015a} are not specified in the table because they adopt the Bayesian hierarchical modelling, where the Jeans equation appears as an integrated form without explicit dependence on stellar and anisotropy models. We also note that \citet{Bonnivard2015a} takes the uncertainty coming from the triaxiality into account to estimate $J$-factor uncertainties.}
    \label{tab:works}
\end{table*}

\begin{figure*}
    \centering
    \includegraphics[width=0.8\hsize]{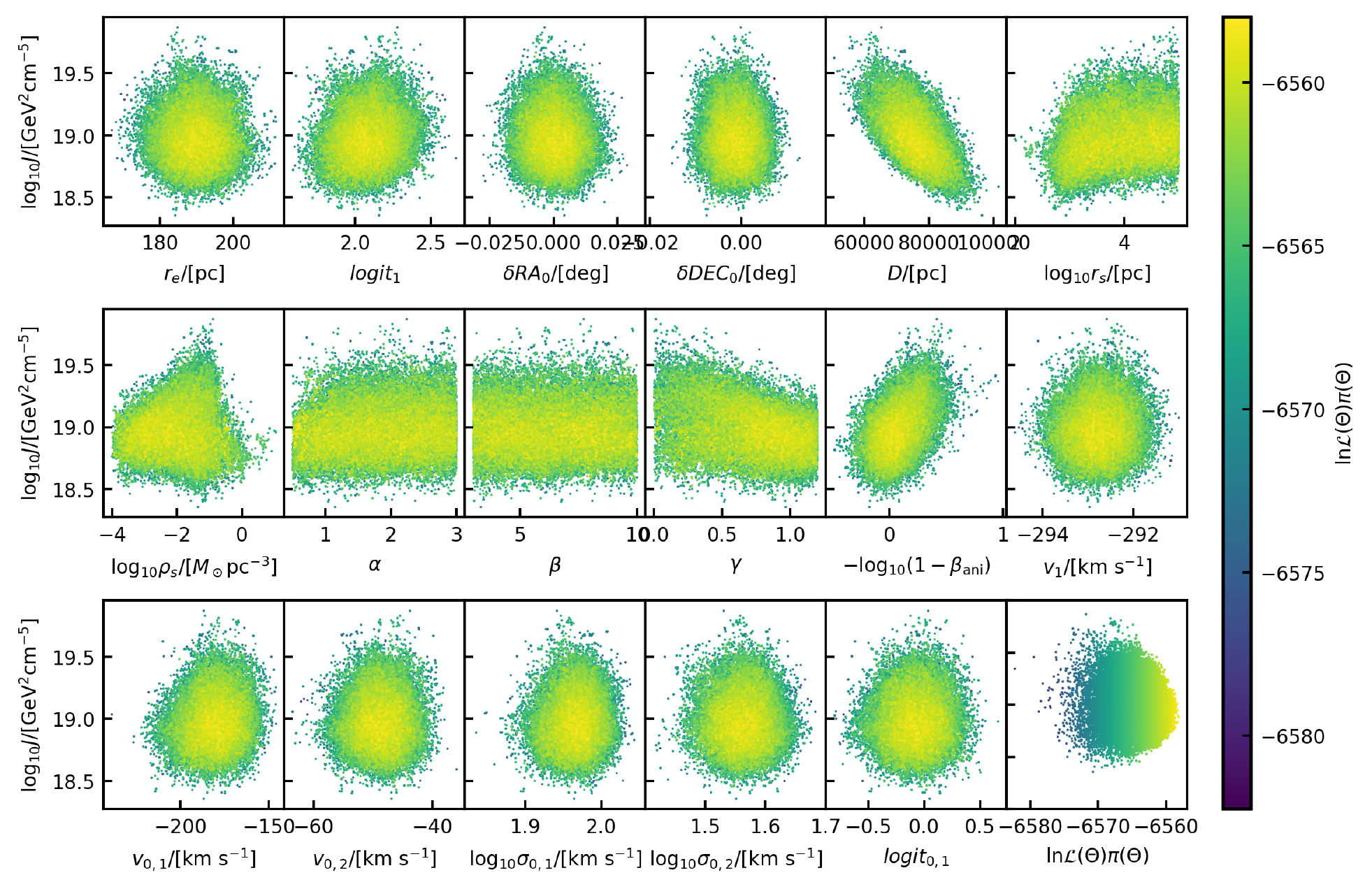}
    \caption{\small Correlation between the $J$-factor PDF and various model parameters for Draco dSph.} 
    \label{fig:corr_J}
\end{figure*}

Finally, \Cref{fig:corr_J,fig:corr_J_scl,fig:corr_J_umi} show various correlations between the estimated the J-factor and the model parameters $\Theta_\spec$ for Draco, Sculptor and Ursa Minor dSphs, respectively.

\section{Discussion}\label{sec:discussion}
We consider implication of our result presented in the previous section, and discuss the impact of the contamination effect on $J$-factor estimation. First, we discuss the advantage or our mixture model to estimate proper DM density distributions as well as $J$-factors of dSphs. Next, we compare our $J$-factor estimation with those of other studies, and figure out common features among them. Finally, we address characteristics of the $J$-factor for each dSph. 

\subsection{Common features}\label{sec:common_features}
The result in Fig.\,\ref{fig:specfit} suggests the importance of properly taking an uncertainty from membership probability into account. It shows that line-of-sight velocity dispersion predicted by our method has a large uncertainty at the outer region, where the most observed stars are indeed from foreground stars. In terms of the Bayesian statistical estimation, data points at the region have more importance than those at regions with many data points. In the conventional analysis, many foreground stars at the outer region are likely to be misidentified as "member" stars even if we impose a rigorous membership selection. Hence, parameter estimation based on the membership selection is often affected by the contamination effect, and it induces an additional systematical uncertainty. In contrast, the mixture model can deal with this uncertainty by the statistical analysis in a straightforward way, because the model includes the contamination effect in the definition of the likelihood function.

On the other hand, \cref{fig:corr_J,fig:corr_J_scl,fig:corr_J_umi} show that estimated $J$-factors are correlated with the distance parameter $D$. The correlation between the $J$-factor and the distance $D$ is fitted by the linear regression, $\log_{10}(J/\text{GeV}^2\text{cm}^{-5}) = a\log_{10}(D/\text{pc}) + b$, with regression coefficients being $a = -3.23$ and $b = 34.73$. As discussed in Appendix\,\ref{sec:distance_err}, the coefficient $a$ takes a value of $a\sim3$, where the small difference from the exact expectation ($a=3$) comes from other uncertainties such as the contamination effect. Indeed, the coefficient is closer to the expected value for the Sculptor case ($a=-2.93$) due to a lower contamination ($\mathrm{logit}_1 \sim 4.2$ or $s_{\Rhalf} \sim $0.99). The uncertainty of the distance between dSph and solar system causes a non-negligible uncertainty on  $J$-factor estimation, especially for the case of dSph with low contamination such as Sculptor.

\cref{fig:corr_J,fig:corr_J_scl,fig:corr_J_umi} also show that the $J$-factor can be correlated with  the DM profile parameter $\gamma$ (inner slope). It indicates that the analysis adopting the generalised NFW profile (or some other analyses that allow the DM profile having an enough freedom to change the inner slope) is important rather than analyses with the profile having a fixed inner slope from the first beginning.

Fig.\,\ref{fig:Jresults} shows that $J$-factors obtained in our analysis are more or less consistent with those obtained by other studies, however their means and errors are slightly different from our result. These differences can be used to discuss validity of assumptions made in each study. For instance, the error bars of \citet{Hayashi2016} are larger than other studies, and it suggests that the uncertainty from the axisymmetricity of dSph is underestimated by assuming the sphericity of dSph. Indeed, (log-)Bayes factors complied by \citet{Hayashi2016} take large values ($> 10$), which is larger than those obtained by analyses assuming the sphericity.

\subsection{Draco}
The value of the J-factor for the Draco dSph reported by \citet{Geringer-Sameth2015} is slightly smaller than ours, and it must be from the contamination effect, because their analysis is almost the same as ours except the treatment of the contamination (and some prior setups). Indeed, in contrast with the fact that our "member" stars in Fig.\,\ref{fig:specfit} are located within $R\lesssim500\ \text{pc}$, their "member" stars shown in Figure 1 of \citet{Geringer-Sameth2015} are distributed as far as $R\sim 1000\ \text{pc}$. The stars and their velocity dispersion with a somewhat suppressed value at the outer region are expected to be from the flatting bias of their EM algorithm. This suppressed dispersion profile is then fitted by a more radial anisotropy.\footnote{\citet{Geringer-Sameth2015} reported $-\log_{10}(1-\betaani) = 0.54_{-0.29}^{+0.27}$, while our result is more isotropic, $-\log_{10}(1-\betaani) = 0.06_{-0.14}^{+0.16}$.}  The suppressed profile, on the other hand, also requires a more concentrated DM profile (smaller $r_s$ with $\rho(r)$ being enough large at the inner region). As a result, the $J$-factor in their analysis is predicted to be smaller than what we have obtained. \citet{Geringer-Sameth2015} reported less $J$-factor error than ours, and this is partially because they utilise information of metallicity for the stellar membership calculation.

\subsection{Sculptor}
The value of the $J$-factor for Sculptor dSph is not sizeably different from each other. This is because the contamination rate of the Sculptor is quite small as mentioned in \cref{sec:common_features}, and the $P_M$ cut procedure works very well. \Cref{fig:Jresults} also shows that the error bar of \citet{Geringer-Sameth2015} is smaller than ours, which is partially because they imposed additional kinematical and cosmological constraints on DM profiles, namely, the criteria on truncation radius and central density. It is also worth notifying that the $J$-factor of Sculptor is significantly dependent on its distance to the solar system due to the low contamination uncertainty, as mentioned in \cref{sec:common_features}. This strong correlation suggests that the major uncertainty of the $J$-factor is from the uncertainty of the distance $D$, and it means that further study of the distance determination is required to obtain a more precise $J$-factor for the Sculptor dSphs.

\subsection{Ursa Minor}
In contrast to the Draco case, \citet{Geringer-Sameth2015} obtained a slightly larger value of the $J$-factor than ours.\footnote{\citet{Bonnivard2015a} also obtained a significantly larger $J$-factor than ours. This is because that they adopted a larger truncation radius than the outermost radius, which is motivated from the tidal radius of the Ursa Minor.} This fact is expected to be from the contamination as in the case of the Draco, though it affects the dispersion profile in a different way. Our dispersion profile has its maximum at $R\simeq 1000\ \text{pc}$, while the maximum of their profile is at $R\simeq 200\ \text{pc}$. The maximum at smaller $R$ in their analysis requires more compact DM profile (smaller $r_s$ and larger $\rho_s$) than ours. On the other hand, the compact DM profile also induces the increase of the profile at inner region, and it must be compensated by more tangential anisotropy.\footnote{Indeed, \citet{Geringer-Sameth2015} obtained tangential anisotropy, $-\log_{10}(1-\betaani) = -0.47_{-0.32}^{+0.28}$, and compact DM profile, $\log_{10}(\rho_s/M_\odot\ \text{pc}^3) = {-0.50}_{-0.64}^{+0.60}$ \& $\log_{10}(r_s/\text{pc}) = 2.60_{-0.38}^{+0.40}$, while we obtained less tangential anisotropy, $-\log_{10}(1-\betaani) = -0.20_{-0.24}^{+0.18}$, and more diluted DM profile, $\log_{10}(\rho_s/M_\odot\ \text{pc}^3) = {-1.55}_{-1.07}^{+0.69}$ \& $\log_{10}(r_s/\text{pc}) = 3.42_{-0.50}^{+0.91}$.} As a result, the $J$-factor value of \citet{Geringer-Sameth2015} becomes larger than our result. Here, it is worth notifying that the compact DM profile contributes to the $J$-factor values conversely compared to the Draco case. This difference comes from the fact that the flattened dispersion profile of the Draco requires the compact DM profile and radial anisotropy, while that of the Sculptor requires the compact DM profile and tangential anisotropy. Moreover, we obtained less $J$-factor error than what \citet{Geringer-Sameth2015} reported, and this is partially because our estimate of $r_s$ is larger than the truncation radius of the $J$-factor, so that the uncertainty of the DM profile does not contribute the $J$-factor uncertainty so much.

\section{Conclusions}
\label{sec:Conclusions}
We have estimated the $J$-factors  of Draco, Sculptor and Ursa Minor dSphs, which are known to be promising targets for the indirect DM detection utilising various gamma-ray observations. We have adopted a mixture model for member and foreground star distributions based on a conditional likelihood function for a given projected distance from the dSph centre, which is obtained by improving the likelihood function proposed in \KI\ and useful to remove sampling bias that we often suffer in spectroscopic observation. 

We introduced a new parameter $\Odds(\Rhalf)$ concerning the membership ratio, which allows us to have a well-determined prior function of the parameter from photometric observation. $J$-factors obtained in our analysis are consistent with those of previous studies, but we saw some small differences among the studies at the same time, which is in particular apparent for higher-contaminated dSphs such as Draco and Ursa Minor. Moreover, we found that the uncertainty of the distance measurement gives a sizeable uncertainty on $J$-factor estimation in some cases. It is thus important to determine the distance accurately to estimate $J$-factor precisely.

Spectroscopic observation with a large field of view in the near future such as the Prime Focus Spectrograph (PFS) will enable us to observe thousands of stars simultaneously and to estimate $J$-factors for various dSphs more precisely. Proper treatment of the foreground contamination hence becomes more and more important, for the number of contaminated foreground stars increases in the data set. In particular, we can expect that the method developed in this paper will be a powerful tool to estimate $J$-factors of ultra-faints dSphs in near future, as they suffer the contamination more seriously.  

\section*{Acknowledgements}
We would like to give special thanks to Koji Ichikawa for his recent researches, as well as the kindness to give us his old computational resources. This research made use of Astropy,\footnote{http://www.astropy.org} a community-developed core Python package for Astronomy \citep{AstropyPaperI, AstropyPaperII}. This work was supported by JSPS KAKENHI Grant Numbers, 18H04359 \& 18J00277 for KH,  15H05889 \& 16H03991 \& 18H05542 for MI, 17K14249 for MNI, 19H05810 \& 16H02176 for SM, 17H02878 for MI \& SM, and 18J21186 for SH. Finally, Kavli IPMU is supported by World Premier International Research Centre Initiative (WPI), MEXT, Japan.

%%%%%%%%%%%%%%%%%%%%%%%%
%%%%%%%% REFERENCES %%%%%%%%
%%%%%%%%%%%%%%%%%%%%%%%%
\bibliographystyle{mnras}
\bibliography{ref_cited} % if your bibtex file is called example.bib

\begin{thebibliography}{}
\makeatletter
\relax
\def\mn@urlcharsother{\let\do\@makeother \do\$\do\&\do\#\do\^\do\_\do\%\do\~}
\def\mn@doi{\begingroup\mn@urlcharsother \@ifnextchar [ {\mn@doi@}
  {\mn@doi@[]}}
\def\mn@doi@[#1]#2{\def\@tempa{#1}\ifx\@tempa\@empty \href
  {http://dx.doi.org/#2} {doi:#2}\else \href {http://dx.doi.org/#2} {#1}\fi
  \endgroup}
\def\mn@eprint#1#2{\mn@eprint@#1:#2::\@nil}
\def\mn@eprint@arXiv#1{\href {http://arxiv.org/abs/#1} {{\tt arXiv:#1}}}
\def\mn@eprint@dblp#1{\href {http://dblp.uni-trier.de/rec/bibtex/#1.xml}
  {dblp:#1}}
\def\mn@eprint@#1:#2:#3:#4\@nil{\def\@tempa {#1}\def\@tempb {#2}\def\@tempc
  {#3}\ifx \@tempc \@empty \let \@tempc \@tempb \let \@tempb \@tempa \fi \ifx
  \@tempb \@empty \def\@tempb {arXiv}\fi \@ifundefined
  {mn@eprint@\@tempb}{\@tempb:\@tempc}{\expandafter \expandafter \csname
  mn@eprint@\@tempb\endcsname \expandafter{\@tempc}}}

\bibitem[\protect\citeauthoryear{Abbott et~al.,}{Abbott
  et~al.}{2018}]{Abbott2018}
Abbott T. M.~C.,  et~al., 2018, \mn@doi [ApJS] {10.3847/1538-4365/aae9f0}, 239,
  18

\bibitem[\protect\citeauthoryear{Abolfathi et~al.,}{Abolfathi
  et~al.}{2018}]{Abolfathi2018}
Abolfathi B.,  et~al., 2018, \mn@doi [ApJS] {10.3847/1538-4365/aa9e8a}, 235, 42

\bibitem[\protect\citeauthoryear{Ackermann et~al.,}{Ackermann
  et~al.}{2015}]{Ackermann2015a}
Ackermann M.,  et~al., 2015, \mn@doi [Phys. Rev. Lett.]
  {10.1103/PhysRevLett.115.231301}, 115, 231301

\bibitem[\protect\citeauthoryear{Ade \& {others}}{Ade \&
  {others}}{2016}]{Ade:2015xua}
Ade P. A.~R.,  {others} 2016, \mn@doi [A{\&}A] {10.1051/0004-6361/201525830},
  594, A13

\bibitem[\protect\citeauthoryear{Baes \& Van~Hese}{Baes \&
  Van~Hese}{2007}]{Baes:2007tx}
Baes M.,  Van~Hese E.,  2007, \mn@doi [A{\&}A] {10.1051/0004-6361:20077672},
  471, 419

\bibitem[\protect\citeauthoryear{Bhattacherjee, Ibe, Ichikawa, Matsumoto  \&
  Nishiyama}{Bhattacherjee et~al.}{2014}]{Bhattacherjee2014}
Bhattacherjee B.,  Ibe M.,  Ichikawa K.,  Matsumoto S.,   Nishiyama K.,  2014,
  \mn@doi [J. High Energy Phys.] {10.1007/JHEP07(2014)080}, 2014, 80

\bibitem[\protect\citeauthoryear{Binney \& Tremaine}{Binney \&
  Tremaine}{2008}]{Binney2008}
Binney J.,  Tremaine S.,  2008, {Galactic Dynamics: Second Edition}.
Princeton Univ. Press, Princeton, NJ

\bibitem[\protect\citeauthoryear{Bonnivard et~al.,}{Bonnivard
  et~al.}{2015}]{Bonnivard2015a}
Bonnivard V.,  et~al., 2015, \mn@doi [MNRAS] {10.1093/mnras/stv1601}, 453, 849

\bibitem[\protect\citeauthoryear{Bonnivard, Maurin  \& Walker}{Bonnivard
  et~al.}{2016}]{Bonnivard2016}
Bonnivard V.,  Maurin D.,   Walker M.~G.,  2016, \mn@doi [MNRAS]
  {10.1093/mnras/stw1691}, 462, 223

\bibitem[\protect\citeauthoryear{Bradac et~al.,}{Bradac
  et~al.}{2006}]{Bradac:2006er}
Bradac M.,  et~al., 2006, \mn@doi [ApJ] {10.1086/508601}, 652, 937

\bibitem[\protect\citeauthoryear{Burkert}{Burkert}{1995}]{Burkert1995}
Burkert A.,  1995, \mn@doi [ApJ] {10.1086/309560}, 447, 10

\bibitem[\protect\citeauthoryear{Chambers et~al.,}{Chambers
  et~al.}{2016}]{Chambers2016}
Chambers K.~C.,  et~al., 2016

\bibitem[\protect\citeauthoryear{Cholis \& Salucci}{Cholis \&
  Salucci}{2012}]{Cholis2012}
Cholis I.,  Salucci P.,  2012, \mn@doi [Phys. Rev. D]
  {10.1103/PhysRevD.86.023528}, 86, 023528

\bibitem[\protect\citeauthoryear{Clowe, Bradac, Gonzalez, Markevitch, Randall,
  Jones  \& Zaritsky}{Clowe et~al.}{2006}]{Clowe:2006eq}
Clowe D.,  Bradac M.,  Gonzalez A.~H.,  Markevitch M.,  Randall S.~W.,  Jones
  C.,   Zaritsky D.,  2006, \mn@doi [ApJ] {10.1086/508162}, 648, L109

\bibitem[\protect\citeauthoryear{Coleman, Da~Costa  \& Bland-Hawthorn}{Coleman
  et~al.}{2005}]{Coleman2005b}
Coleman M.~G.,  Da~Costa G.~S.,   Bland-Hawthorn J.,  2005, \mn@doi [AJ]
  {10.1086/432662}, 130, 1065

\bibitem[\protect\citeauthoryear{Dehnen}{Dehnen}{1993}]{Dehnen1993}
Dehnen W.,  1993, \mn@doi [MNRAS] {10.1093/mnras/265.1.250}, 265, 250

\bibitem[\protect\citeauthoryear{Drlica-Wagner et~al.,}{Drlica-Wagner
  et~al.}{2017}]{Drlica-Wagner2017a}
Drlica-Wagner A.,  et~al., 2017, ] {10.3847/1538-4365/aab4f5}

\bibitem[\protect\citeauthoryear{Evans, Ibe, Olive  \& Yanagida}{Evans
  et~al.}{2014}]{Evans2014UniversalityMediation}
Evans J.~L.,  Ibe M.,  Olive K.~A.,   Yanagida T.~T.,  2014, Technical report,
  {Universality in Pure Gravity Mediation}

\bibitem[\protect\citeauthoryear{Foreman-Mackey, Hogg, Lang  \&
  Goodman}{Foreman-Mackey et~al.}{2013}]{Foreman-Mackey2013}
Foreman-Mackey D.,  Hogg D.~W.,  Lang D.,   Goodman J.,  2013, \mn@doi [PASP]
  {10.1086/670067}, 125, 306

\bibitem[\protect\citeauthoryear{Geringer-Sameth, Koushiappas  \&
  Walker}{Geringer-Sameth et~al.}{2015}]{Geringer-Sameth2015}
Geringer-Sameth A.,  Koushiappas S.~M.,   Walker M.,  2015, \mn@doi [ApJ]
  {10.1088/0004-637X/801/2/74}, 801, 74

\bibitem[\protect\citeauthoryear{Hastings}{Hastings}{1970}]{Hastings:1970aa}
Hastings W.~K.,  1970, \mn@doi [Biometrika] {10.1093/biomet/57.1.97}, 57, 97

\bibitem[\protect\citeauthoryear{Hayashi, Ichikawa, Matsumoto, Ibe, Ishigaki
  \& Sugai}{Hayashi et~al.}{2016}]{Hayashi2016}
Hayashi K.,  Ichikawa K.,  Matsumoto S.,  Ibe M.,  Ishigaki M.~N.,   Sugai H.,
  2016, \mn@doi [MNRAS] {10.1093/mnras/stw1457}, 461, 2914

\bibitem[\protect\citeauthoryear{Hernquist}{Hernquist}{1990}]{Hernquist:1990}
Hernquist L.,  1990, \mn@doi [ApJ] {10.1086/168845}, 356, 359

\bibitem[\protect\citeauthoryear{Hisano, Matsumoto, Nagai, Saito  \&
  Senami}{Hisano et~al.}{2007}]{Hisano:2006nn}
Hisano J.,  Matsumoto S.,  Nagai M.,  Saito O.,   Senami M.,  2007, \mn@doi
  [Phys. Lett.] {10.1016/j.physletb.2007.01.012}, B646, 34

\bibitem[\protect\citeauthoryear{Ichikawa, Ishigaki, Matsumoto, Ibe, Sugai,
  Hayashi  \& Horigome}{Ichikawa et~al.}{2017}]{Ichikawa2017}
Ichikawa K.,  Ishigaki M. N. M.~N.,  Matsumoto S.,  Ibe M.,  Sugai H.,  Hayashi
  K.,   Horigome S.-i. S.-i.,  2017, \mn@doi [MNRAS] {10.1093/mnras/stx682},
  468, 2884

\bibitem[\protect\citeauthoryear{Ichikawa et~al.,}{Ichikawa
  et~al.}{2018}]{Ichikawa}
Ichikawa K.,  et~al., 2018, \mn@doi [MNRAS] {10.1093/mnras/sty1387}, 479, 64

\bibitem[\protect\citeauthoryear{Irwin \& Hatzidimitriou}{Irwin \&
  Hatzidimitriou}{1995}]{Irwin1995}
Irwin M.,  Hatzidimitriou D.,  1995, \mn@doi [MNRAS]
  {10.1093/mnras/277.4.1354}, 277, 1354

\bibitem[\protect\citeauthoryear{Jasra, Holmes  \& Stephens}{Jasra
  et~al.}{2005}]{Jasra2005MarkovModeling}
Jasra A.,  Holmes C.~C.,   Stephens D.~A.,  2005, \mn@doi [Statistical Science]
  {10.1214/088342305000000016}, 20, 50

\bibitem[\protect\citeauthoryear{Koch, Kleyna, Wilkinson, Grebel, Gilmore,
  Evans, Wyse  \& Harbeck}{Koch et~al.}{2007}]{Koch:2007ye}
Koch A.,  Kleyna J.~T.,  Wilkinson M.~I.,  Grebel E.~K.,  Gilmore G.~F.,  Evans
  N.~W.,  Wyse R. F.~G.,   Harbeck D.~R.,  2007, \mn@doi [AJ] {10.1086/519380},
  134, 566

\bibitem[\protect\citeauthoryear{Lefranc, Moulin, Panci, Sala  \& Silk}{Lefranc
  et~al.}{2016}]{Lefranc2016}
Lefranc V.,  Moulin E.,  Panci P.,  Sala F.,   Silk J.,  2016, \mn@doi [JCAP]
  {10.1088/1475-7516/2016/09/043}, 2016, 043

\bibitem[\protect\citeauthoryear{Martin, de Jong  \& Rix}{Martin
  et~al.}{2008}]{Martin2008}
Martin N.~F. N.~F.,  de Jong J. T. A. J.~T.~A.,   Rix H.-W. H.-W.,  2008,
  \mn@doi [ApJ] {10.1086/590336}, 684, 1075

\bibitem[\protect\citeauthoryear{Martinez}{Martinez}{2015}]{Martinez:2013els}
Martinez G.~D.,  2015, \mn@doi [MNRAS] {10.1093/mnras/stv942}, 451, 2524

\bibitem[\protect\citeauthoryear{Martinez, Bullock, Kaplinghat, Strigari  \&
  Trotta}{Martinez et~al.}{2009}]{Martinez:2009jh}
Martinez G.~D.,  Bullock J.~S.,  Kaplinghat M.,  Strigari L.~E.,   Trotta R.,
  2009, \mn@doi [JCAP] {10.1088/1475-7516/2009/06/014}, 0906, 14

\bibitem[\protect\citeauthoryear{Martinez, Minor, Bullock, Kaplinghat, Simon
  \& Geha}{Martinez et~al.}{2011}]{Martinez2011}
Martinez G.~D.,  Minor Q.~E.,  Bullock J.,  Kaplinghat M.,  Simon J.~D.,   Geha
  M.,  2011, \mn@doi [ApJ] {10.1088/0004-637X/738/1/55}, 738, 55

\bibitem[\protect\citeauthoryear{Mateo, Olszewski  \& Walker}{Mateo
  et~al.}{2008}]{Mateo2008}
Mateo M.,  Olszewski E.~W.,   Walker M.~G.,  2008, \mn@doi [ApJ]
  {10.1086/522326}, 675, 201

\bibitem[\protect\citeauthoryear{McConnachie}{McConnachie}{2012}]{McConnachie2012a}
McConnachie A.~W.,  2012, \mn@doi [AJ] {10.1088/0004-6256/144/1/4}, 144, 4

\bibitem[\protect\citeauthoryear{McLaughlin}{McLaughlin}{1999}]{McLaughlin:1998sb}
McLaughlin D.~E.,  1999, \mn@doi [ApJ] {10.1086/311860}, 512, L9

\bibitem[\protect\citeauthoryear{Metropolis, Rosenbluth, Rosenbluth, Teller  \&
  Teller}{Metropolis et~al.}{1953}]{Metropolis1953}
Metropolis N.,  Rosenbluth A.~W.,  Rosenbluth M.~N.,  Teller A.~H.,   Teller
  E.,  1953, \mn@doi [J. Chem. Phys.] {10.1063/1.1699114}, 21, 1087

\bibitem[\protect\citeauthoryear{Moroi \& Randall}{Moroi \&
  Randall}{2000}]{Moroi2000}
Moroi T.,  Randall L.,  2000, \mn@doi [Nuclear Phys. B]
  {10.1016/S0550-3213(99)00748-8}, 570, 455

\bibitem[\protect\citeauthoryear{Navarro, Frenk  \& White}{Navarro
  et~al.}{1997}]{Navarro:1997}
Navarro J.~F.,  Frenk C.~S.,   White S. D.~M.,  1997, \mn@doi [ApJ]
  {10.1086/304888}, 490, 493

\bibitem[\protect\citeauthoryear{Plummer}{Plummer}{1911}]{Plummer1911}
Plummer H.~C.~C.,  1911, \mn@doi [MNRAS] {10.1093/mnras/71.5.460}, 71, 460

\bibitem[\protect\citeauthoryear{Price-Whelan et~al.,}{Price-Whelan
  et~al.}{2018}]{AstropyPaperII}
Price-Whelan A.~M.,  et~al., 2018, \mn@doi [AJ] {10.3847/1538-3881/aabc4f},
  156, 123

\bibitem[\protect\citeauthoryear{Robin, Reyle, Derriere, Picaud, Reyl{\'{e}},
  Derri{\`{e}}re  \& Picaud}{Robin et~al.}{2003}]{Robin:2003}
Robin A.~C.,  Reyle C.,  Derriere S.,  Picaud S.,  Reyl{\'{e}} C.,
  Derri{\`{e}}re S.,   Picaud S.,  2003, \mn@doi [A{\&}A]
  {10.1051/0004-6361:20040968}, 409, 523

\bibitem[\protect\citeauthoryear{Robitaille et~al.,}{Robitaille
  et~al.}{2013}]{AstropyPaperI}
Robitaille T.~P.,  et~al., 2013, \mn@doi [A{\&}A]
  {10.1051/0004-6361/201322068}, 558, A33

\bibitem[\protect\citeauthoryear{Rubin, Thonnard, Ford  \& Ford~Jr.}{Rubin
  et~al.}{1978}]{Rubin1978}
Rubin V.~C.~C.,  Thonnard N.,  Ford W.~K. J.,   Ford~Jr. W.~K.,  1978, \mn@doi
  [ApJ] {10.1086/182804}, 225, L107

\bibitem[\protect\citeauthoryear{Rubin, Ford, Thonnard  \& Ford}{Rubin
  et~al.}{1980}]{Rubin1980a}
Rubin V.~C. V.~C.,  Ford W.~K.~J.,  Thonnard N.,   Ford W.~K. J.,  1980,
  \mn@doi [ApJ] {10.1086/158003}, 238, 471

\bibitem[\protect\citeauthoryear{Schwarz}{Schwarz}{1978}]{Schwarz1978}
Schwarz G.,  1978, \mn@doi [Ann. Statistics] {10.1214/aos/1176344136}, 6, 461

\bibitem[\protect\citeauthoryear{S{\'{e}}gall, Ibata, Irwin, Martin  \&
  Chapman}{S{\'{e}}gall et~al.}{2007}]{Segall2007}
S{\'{e}}gall M.,  Ibata R.~A.,  Irwin M.~J.,  Martin N.~F.,   Chapman S.,
  2007, \mn@doi [MNRAS] {10.1111/j.1365-2966.2006.11356.x}, 375, 831

\bibitem[\protect\citeauthoryear{Simon \& Geha}{Simon \&
  Geha}{2007}]{Simon2007}
Simon J.~D.,  Geha M.,  2007, \mn@doi [ApJ] {10.1086/521816}, 670, 313

\bibitem[\protect\citeauthoryear{Spencer, Mateo, Olszewski, Walker, McConnachie
   \& Kirby}{Spencer et~al.}{2018}]{Spencer2018}
Spencer M.~E.,  Mateo M.,  Olszewski E.~W.,  Walker M.~G.,  McConnachie A.~W.,
   Kirby E.~N.,  2018, \mn@doi [AJ] {10.3847/1538-3881/aae3e4}, 156, 257

\bibitem[\protect\citeauthoryear{Sugai et~al.,}{Sugai
  et~al.}{2015}]{2015JATIS...1c5001S}
Sugai H.,  et~al., 2015, \mn@doi [J. Astron. Telesc. Instrum. Syst.]
  {10.1117/1.JATIS.1.3.035001}, 1, 035001

\bibitem[\protect\citeauthoryear{Takada et~al.,}{Takada
  et~al.}{2014}]{2014PASJ...66R...1T}
Takada M.,  et~al., 2014, \mn@doi [PASJ] {10.1093/pasj/pst019}, 66, R1

\bibitem[\protect\citeauthoryear{Tamura et~al.,}{Tamura
  et~al.}{2016}]{Tamura:2016wsg}
Tamura N.,  et~al., 2016, in Evans C.~J.,  Simard L.,   Takami H.,  eds,  Vol.
  9908, Proc. SPIE Int. Soc. Opt. Eng.. p. 99081M, \mn@doi{10.1117/12.2232103},
  \url
  {http://proceedings.spiedigitallibrary.org/proceeding.aspx?doi=10.1117/12.2232103}

\bibitem[\protect\citeauthoryear{Ullio \& Valli}{Ullio \&
  Valli}{2016}]{Ullio2016}
Ullio P.,  Valli M.,  2016, \mn@doi [JCAP] {10.1088/1475-7516/2016/07/025},
  2016, 025

\bibitem[\protect\citeauthoryear{Walker, Mateo, Olszewski, Sen  \&
  Woodroofe}{Walker et~al.}{2009}]{Walker2009d}
Walker M.~G.,  Mateo M.,  Olszewski E.~W.,  Sen B.,   Woodroofe M.,  2009,
  \mn@doi [AJ] {10.1088/0004-6256/137/2/3109}, 137, 3109

\bibitem[\protect\citeauthoryear{Walker, Olszewski  \& Mateo}{Walker
  et~al.}{2015}]{Walker2015a}
Walker M.~G. M.~G.,  Olszewski E. W. E.~W.,   Mateo M.,  2015, \mn@doi [MNRAS]
  {10.1093/mnras/stv099}, 448, 2717

\bibitem[\protect\citeauthoryear{Watanabe}{Watanabe}{2012}]{Watanabe2012}
Watanabe S.,  2012, Journal of Machine Learning Research, 14, 867

\bibitem[\protect\citeauthoryear{Zhao}{Zhao}{1996}]{Zhao1996}
Zhao H.,  1996, \mn@doi [MNRAS] {10.1093/mnras/278.2.488}, 278, 488

\bibitem[\protect\citeauthoryear{Zwicky}{Zwicky}{1933}]{Zwicky1933a}
Zwicky F.,  1933, \mn@doi [Helvetica Physica Acta] {10.5169/seals-110267}, 6,
  110

\makeatother
\end{thebibliography}

%%%%%%%%%%%%%%%%%%%%%%%%
%%%%%%%% APPENDICES %%%%%%%%
%%%%%%%%%%%%%%%%%%%%%%%%
\appendix

\section{Improving KI17 likelihood}
\label{App:KI17tomodKI17} % In order to avoid warning of the hyperref package, here we use KI17 instead of \KI.

\begin{table}
    \centering
    \begin{tabular}{c|p{5cm}}
        \hline
        Notation & Description \\
        \hline
        $p(v,R)$ & Simultaneous probability to find a star at the radius $R$ and at the velocity $v$.\\
        $p(v,R|M)$ & The same as above but for member stars ($M=1$) and foreground stars ($M=0$). \\
        $p(R|M)$ & Normalised stellar density profile for member and foreground stars ($\int\dd{R} p(R|M) = 1$). \\
        $p(M|R)$ & Local membership probability at radius $R$. \\
        $p(M)$ & Global membership probability. It depends on the CMD cut criteria and RoI selections. \\
        \hline
    \end{tabular}
    \caption{Probabilities and probability density functions used in our likelihood function. Note that we always put random variables on the left-hand-side of the bracket, $(~|~)$, while conditions are put on the right-hand-side.}
    \label{tab:notation}
\end{table}

We derive our likelihood function from that of \KI\,in which the likelihood function for the spectroscopic data $\calL$ is defined as follows:
\begin{equation*}
    \calL = \prod_i \qty[s f_1(v_i,R_i)  + (1-s)f_0(v_i,R_i)]\ .
\end{equation*}
This is based on the probability density function $p(\mathbf{v},\mathbf{R}) \equiv \prod_i \sum_{M_i=1,0} p(M_i) p(v_i,R_i|M_i)$, where $p(M=1) = s$ and $p(v,R|M) = f_M(v,R)$.\footnote{We summarise definitions of the probabilities in Table \ref{tab:notation}.} In order to obtain a likelihood which is free from sampling bias, we use the conditional probability $p(\mathbf{v}|\mathbf{R})$ instead of $p(\mathbf{v},\mathbf{R})$ as a new likelihood function as follows:
\begin{align}
    p(\mathbf{v}|\mathbf{R}) 
    %&= \frac{\sum_{M_i} p(M=M_i)p(\mathbf{v},\mathbf{R}|M_i,\Theta)}{\int_{-\infty}^{\infty}\dd{\mathbf{v}}\sum_{M_i} p(M=M_i)p(\mathbf{v},\mathbf{R}|M_i,\Theta)}\\
    &= \prod_i \sum_{M_i} p(M=M_i|R_i)p(v_i|R_i,M_i) \\
    &\equiv \prod_i \qty[s(R_i) \calG_1(v_i) + [1-s(R_i)] \calG_0(v_i)]\ ,
\end{align}
where the local membership probability $s(R)$ is given by the formula,
\begin{align}
    s(R) &\equiv p(M=1|R) = \frac{p(M=1)\int\dd{v}p(v,R|M=1)}{\sum_M p(M)\int\dd{v}p(v,R|M)} \\
    &= \qty[1+\qty(\frac{s\int\dd{v}f_1(v,R)}{(1-s)\int\dd{v}f_0(v,R)})^{-1}]^{-1}.
\end{align}
Here, it is worth pointing out that the integral $\int\dd{v}f_M(v,R)$ are proportional to the projected stellar density function $\Sigma_M(R)$ when $\int\dd{v} = \int_{-\infty}^{\infty}\dd{v}$, because we assume that the distribution function $f_M(v,R)$ has the velocity dependence only through the Gaussian function, whose integral is given by $\int_{-\infty}^{\infty}\dd{v}\calG(v;v_M,\sigma_M(R)) = 1$.\footnote{In the case of the finite-ranged integration, $\int\dd{v}f_M(v,R)$ gives another $R$-dependent factor, which comes from $\int\dd{v}\calG(v;v_M,\sigma_M(R))$. } In this case, $s_R$ is simplified as follows:
\begin{equation}
    s(R) = \qty[1+\frac{1}{\Odds(\Rhalf)}\frac{\Sigma_1(\Rhalf)/\Sigma_1(R)}{\Sigma_0(\Rhalf)/\Sigma_0(R)}]^{-1}\ ,
\end{equation}
where the ratio $\Sigma_0(R)/\Sigma_0(\Rhalf)$ is always one in our analysis, because we have assumed uniform stellar density for foreground stars. Then, the local odds $\Odds(\Rhalf)$ is given as follows:
\begin{align}
    \Odds(\Rhalf) 
    &\equiv \frac{p(M=1|R=\Rhalf)}{p(M=0|R=\Rhalf)} \\
    &= \frac{s}{1-s}\frac{\int\dd{v}f_1(v,\Rhalf)}{\int\dd{v}f_0(v,\Rhalf)}\ .
\end{align}
Hence, we adopt $\Odds(\Rhalf)$ as an alternative parameter of $s$. The advantage of this parametrization is that we can define the membership probability of a specific dSph without sampling bias, allowing us to utilise several data sets from different observations, for example, the photometry data and the spectroscopy data in this work.\footnote{Moreover, the use of $s$ causes another trouble: when we observe all stars from the dSph centre to the radius $R$ which is much away from the centre, $s$ becomes almost 0, because the number of member stars are finite at $R\to\infty$, while that of contaminating foreground stars is proportional to $R^2$.}

\section{Distance dependence of $J$-factor}\label{sec:distance_err}
$J$-factor depends strongly on the distance to a dSph. \citet{Ullio2016} shows that, for a given distance $D$, a spherical dSph with the DM density profile $\rho_\text{DM}(r)$ has the following dependence:
\begin{equation}
    J_D \simeq \frac{4\pi}{D^2}\int_0^{r_\text{max}}\dd{r}r^2 \rho_\text{DM}^2(r)\ .
\end{equation}
The DM density profile $\rho_\text{DM}(r)$ also depends implicitly on $D$ due to the invariance of the observed line-of-sight velocity dispersion $\sigma^2_{\Theta}(R)$ ($\Theta$: parameters). Namely, for a distance $D' = kD$, a observed radius of a star $R$ is also scaled by the factor $k$ because of $R = D\sin\theta$, and the re-scaled line-of-sight velocity dispersion $\sigma^2_{\Theta'}(R')$ with different parameters $\Theta'\ni D'$ and a radius $R' = kR$ must be equal to the original one as $\sigma^2_{\Theta}(R)= \sigma^2_{\Theta'}(R')$. Here,
\begin{align}
    & \Theta = \qty{\rho_s,\, r_s,\, \Rhalf,\, D}\ , \\ 
    & \Theta' = {\qty{\rho'_s,\, r'_s,\, \Rhalf',\, D'}} = {\qty{k^{-2}\rho_s,\, k r_s,\, k\Rhalf,\, kD}}\ .
\end{align}
This relationship can be confirmed as follows: The stellar number densities and the DM density profile are in general defined as
\begin{align}
    \Sigma_1(R) &= R_e^{-2} f_1(R/R_e)\ , \\
    \nu_1(r) &= R_e^{-3} f_2(r/R_e)\ , \\
    \rho_\text{DM}(r) &= \rho_s f_3 (r/r_s)\ , \\
    M(r) &= \rho_s r_s^3 f_4(r/r_s)\ .
\end{align}
Substituting them into equations (\ref{eq:sigmalos}) and (\ref{eq:sigmar}) leads to
\begin{align}
    \sigma^2_{\Theta'}(R') 
    &= \frac{2}{k^{-2}\Sigma_1(R)}\int_{kR}^\infty\frac{\dd{r}}{\sqrt{1-(kR)^2/r^2}} \qty(1-\beta_\text{ani}\frac{(kR)^2}{r^2}) \notag\\
    &\qq{} \times\int_r^\infty k^{-3}\nu_1(r')\qty(\frac{r'}{r})^{2\beta_\text{ani}}\frac{k^{-2+3}G M(r')}{{r'}^2} \dd{r'} \\
    &= \frac{2}{k^{-2}\Sigma_1(R)}\int_{R}^\infty\frac{k\dd{r}}{\sqrt{1-R^2/r^2}} \qty(1-\beta_\text{ani}\frac{R^2}{r^2}) \notag\\
    &\qq{} \times\int_r^\infty k^{-3}\nu_1(r')\qty(\frac{r'}{r})^{2\beta_\text{ani}}\frac{k^{-2+3}G M(r')}{k^{2}{r'}^2} k\dd{r'} \\
    &= \sigma^2_{\Theta}(R)\ .
\end{align}
It means that our likelihood function has a degeneracy, and we obtain the $J$-factor for a different input distance $D'=kD$ as 
\begin{align}
    J_{D'} 
    &\simeq \frac{4\pi}{k^2D^2}\int_0^{kr_\text{max}}\dd{r}r^2 \eval{\rho_\text{DM}^2(kr)}_{r_s = k r_s}\ \\
    &= \frac{4\pi}{k^2D^2}\int_0^{r_\text{max}}k\dd{r}k^2r^2 k^{-2\times2}\rho_\text{DM}^2(r) = k^{-3} J_{D}\ .
\end{align}
As a result, estimation on the $J$-factor is non-negligibly affected by the uncertainty of the distances $D$. For instance, ten per cent error of the distance $k = 1.0_{-0.1}^{+0.1}$ gives $\log_{10}J_D' = \qty(\log_{10}J_D)_{-0.14}^{+0.12}$.

\section{Widely applicable Bayesian information criterion}\label{sec:wbic}
Due to the difficulty of the multidimensional integration appeared for the statistical evidence, several approximations for the evidence have been developed. The Bayesian Information Criteria (BIC) \citep{Schwarz1978} is a well-known approximation, but it is valid only for Gaussian-like posterior probabilities. The Widely applicable Bayesian Information Criteria (WBIC) \citep{Watanabe2012} is a more generic and easily computable approximation, which is defined by
\begin{equation}
    \text{WBIC} = \frac{\int\dd{\Theta}\qty(-\ln\calL(\Theta)) \calL(\Theta)^\beta \pi(\Theta)}{\int\dd{\Theta}\calL(\Theta)^\beta \pi(\Theta)}\ .
\end{equation}
Here $\beta$ is called the inverse temperature, with is given by $\beta = 1/\ln N$ with $N$ being the number of samples. We can easily calculate the WBIC by MCMC sampling of $-\calL(\Theta)^\beta\pi(\Theta)$. The WBIC gives a good approximation of the minus log-evidence (or the free energy in the statistics) even for singular statistical models, such as the Gaussian Mixture Model (GMM). Our spectroscopic likelihood function contains the GMM, namely the foreground model, hence we adopt the WBIC to approximate the evidence.

%%%%%%%% PDFs %%%%%%%%
\section{PDFs of Sculptor and Ursa Minor}
\label{app:Scl-Umi_figs}
\begin{figure*}
    \centering
    \includegraphics[width=\hsize]{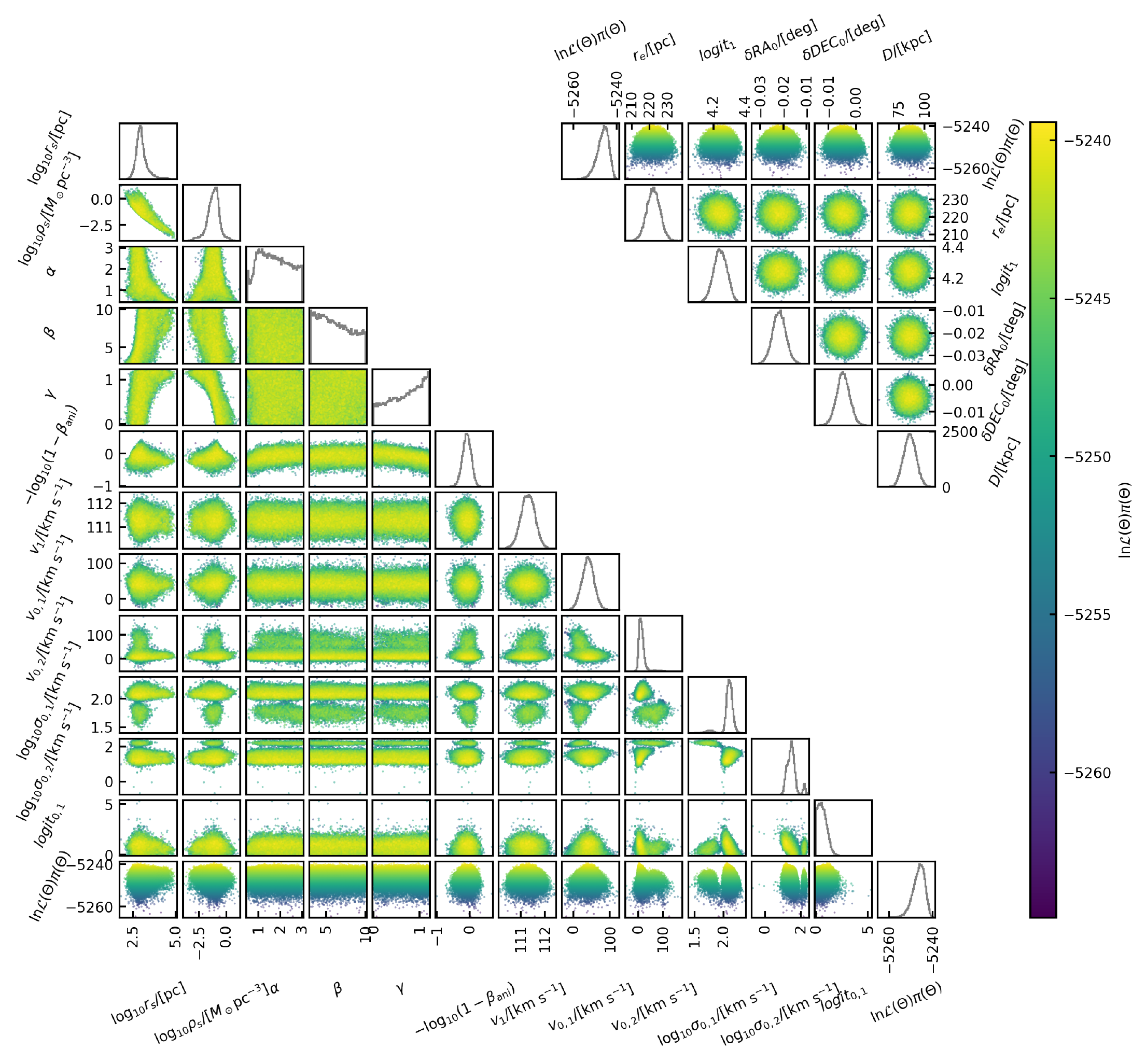}
    \caption{\small Posterior probability density and correlation matrix for the Sculptor. Multimodal distribution that we can see on the PDF is due to the fact that the contamination level of the dSph is very low ($\mathrm{logit}_1 \sim 4.2$ or $s_{\Rhalf} \sim 0.99$), so that the model cannot resolve foreground stars into two Gaussian distributions.}
    \label{fig:scattermatrix_scl}
\end{figure*}

\begin{figure*}
    \centering
    \includegraphics[width=\hsize]{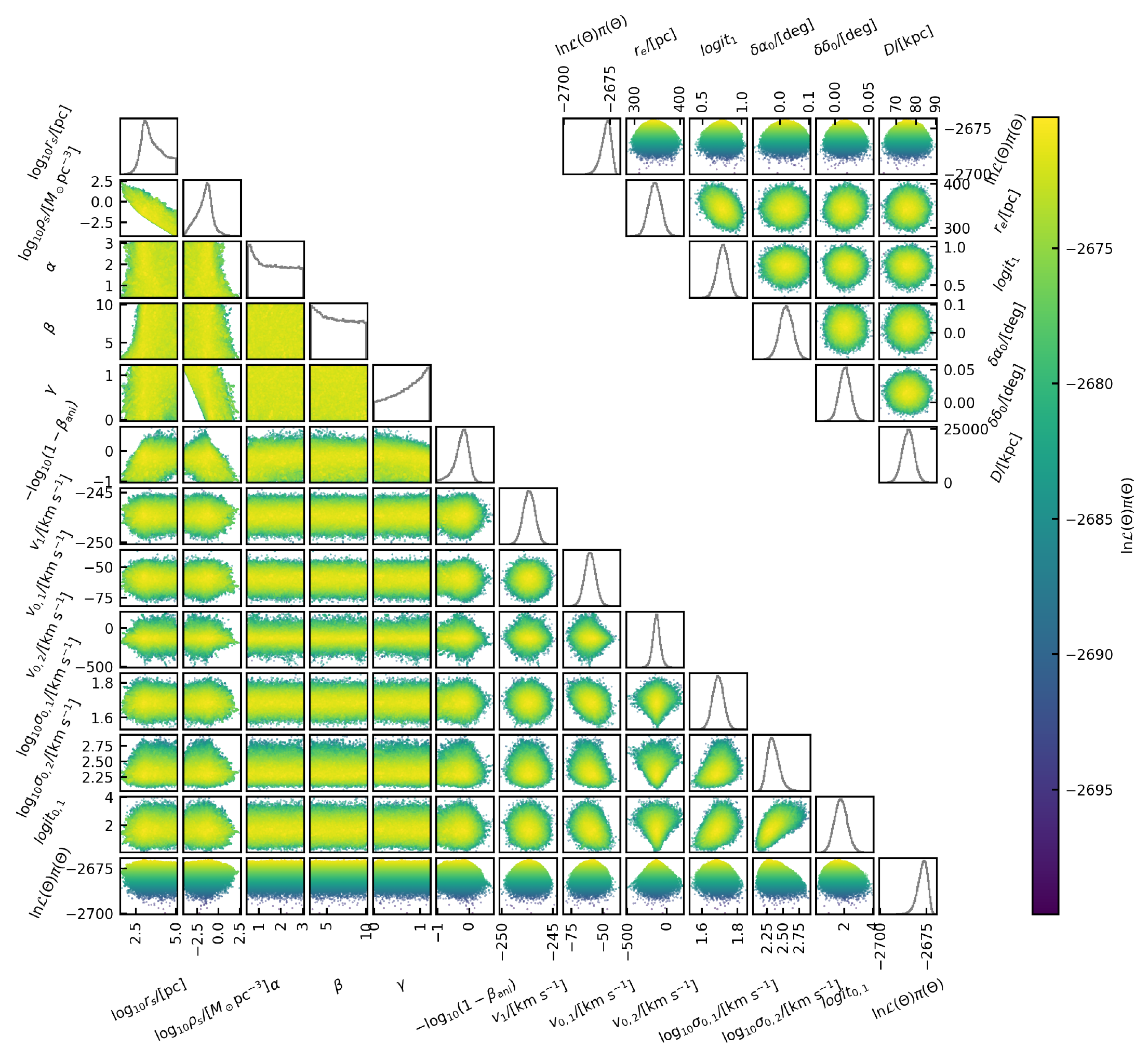}
    \caption{Posterior probability density and correlation matrix for the Ursa Minor.}
    \label{fig:scattermatrix_umi}
\end{figure*}

\begin{figure*}
    \centering
    \includegraphics[width=0.8\hsize]{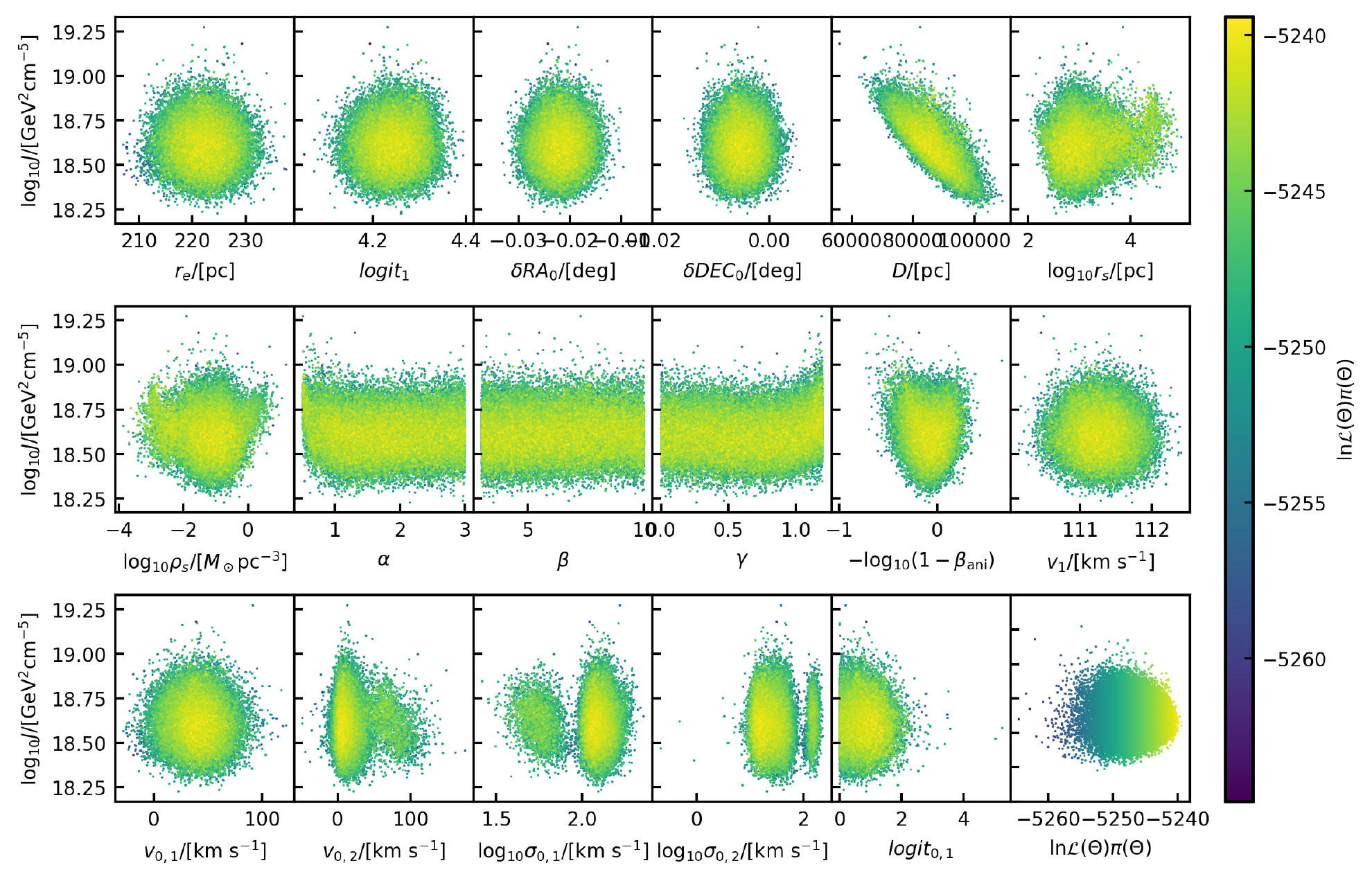}
    \caption{$J$-factor PDFs of the Sculptor dSph with respect to some other parameters.}
    \label{fig:corr_J_scl}
\end{figure*}

\begin{figure*}
    \centering
    \includegraphics[width=0.8\hsize]{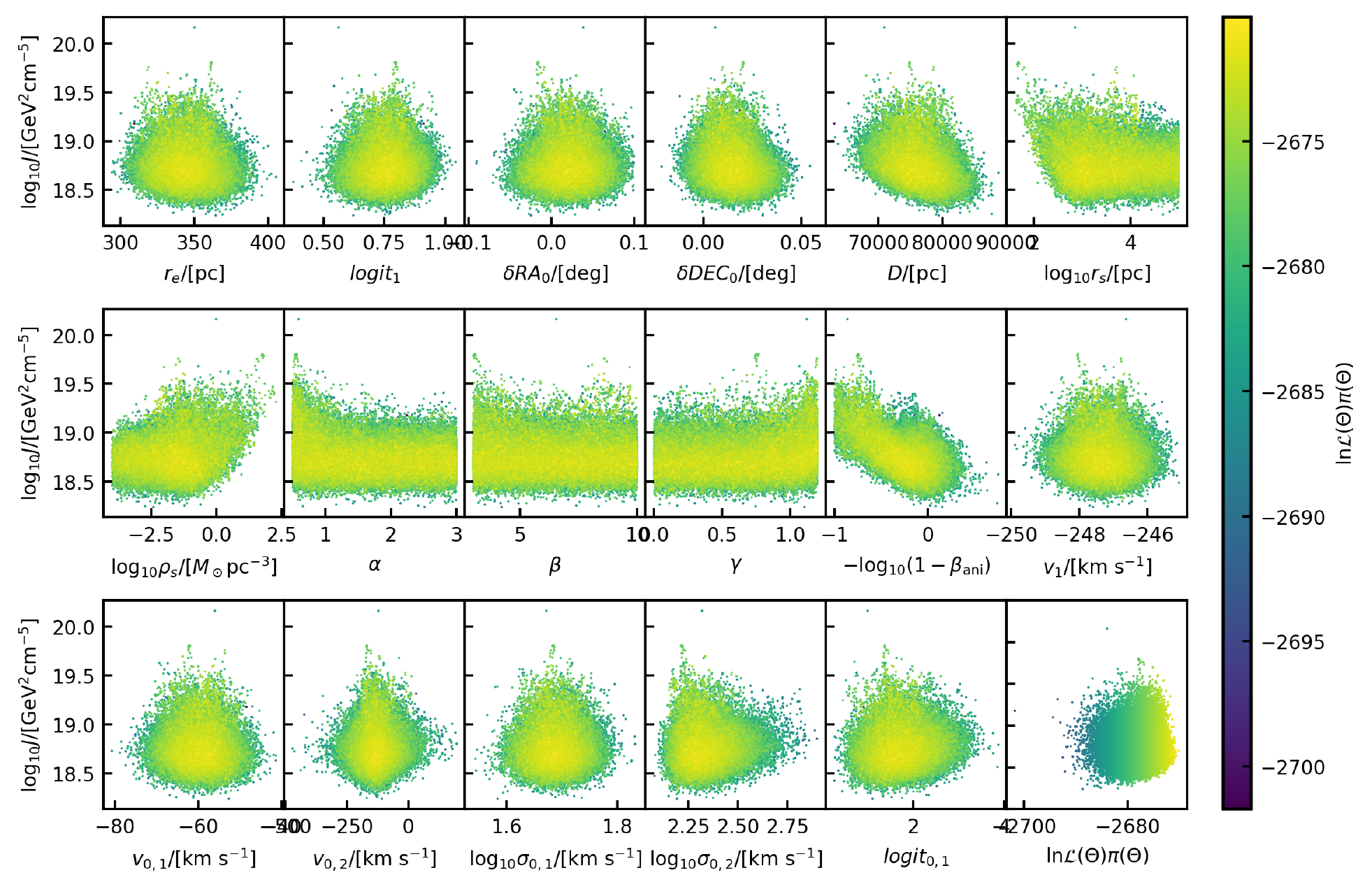}
    \caption{$J$-factor PDFs of the Ursa Minor dSph with respect to some other parameters.}
    \label{fig:corr_J_umi}
\end{figure*}

Here, we show the result of parameter estimation for Sculptor and Ursa Minor: Figs.\,\ref{fig:scattermatrix_scl} and \ref{fig:scattermatrix_umi} are posterior PDFs, while 
Figs.\,\ref{fig:corr_J_scl} and \ref{fig:corr_J_umi} are correlations between their $J$-factors and other parameters.

\section{Truncation}
\begin{figure*}
    \centering
    \includegraphics[width=0.8\hsize]{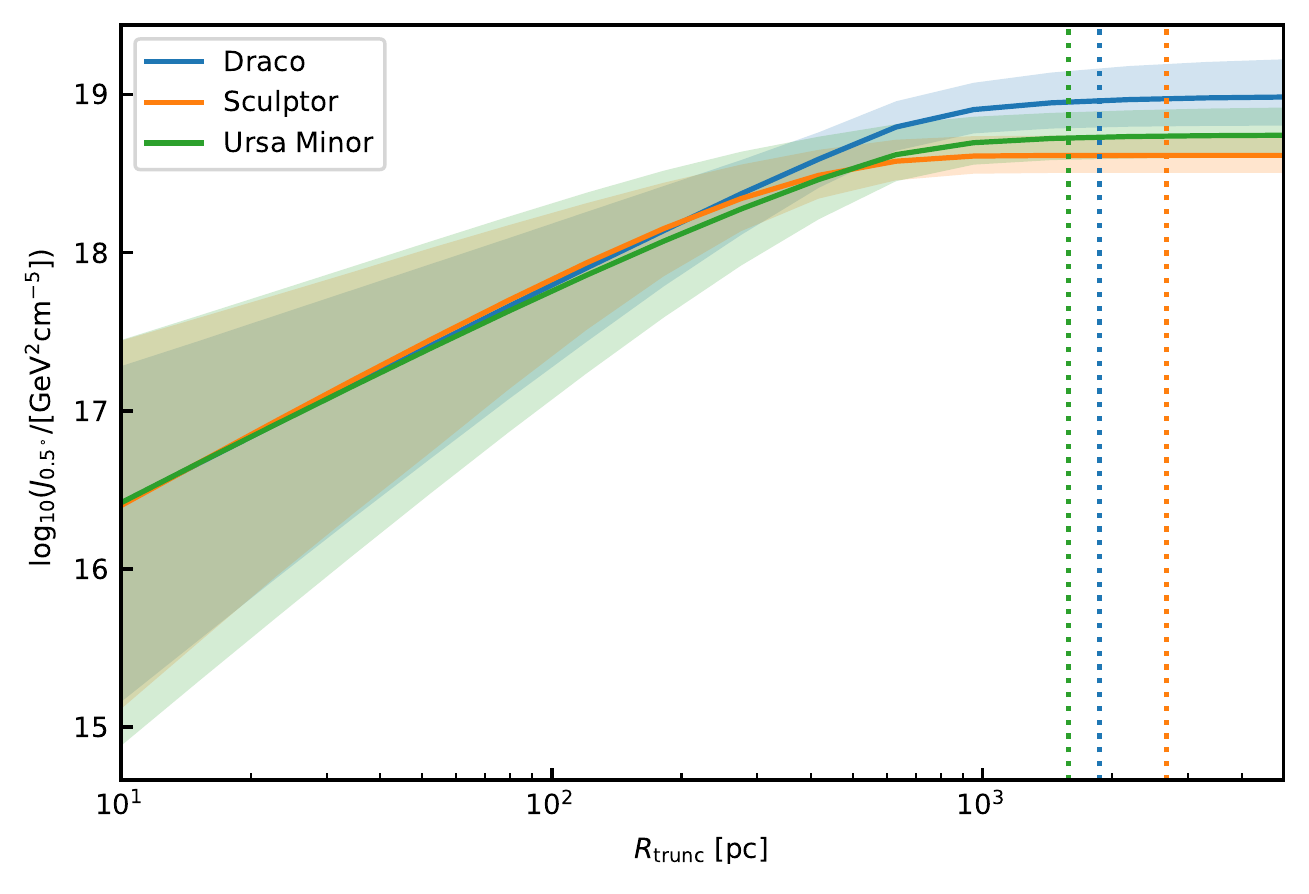}
    \caption{The value of the $J$-factor as a function of the truncation radius $R_\text{trunc}$. Blue, orange, and green lines are the median values of $J$-factor for Draco, Sculptor and Ursa Minor dSphs, respectively. On the other hand, the  shaded area with the same colour code corresponds to the 68\% percentile of the Bayesian credible interval. Vertical dotted lines are the truncation radii that we used in our analysis, and exactly the same as those shown in \Table{DataSource}.}
    \label{fig:truncation}
\end{figure*}

Dependence of the truncation radius $R_\text{trunc}$ on the value of the $J$-factor for each dSph (Draco, Sculptor, Ursa Minor) is shown in Fig.\,\ref{fig:truncation}. The radius that we used in our analysis is shown as a vertical dotted line. It can be seen that the value of the $J$-factor is not very sensitive to the choice of the truncation radius if it is large enough.

%%%%%%%%%%%%%%%%%%%%%%%%%%%%%%%%%%%%%%%%%%%%%%%%%%

% Don't change these lines
\bsp	% typesetting comment
\label{lastpage}
\end{document}